\documentclass[notitlepage]{report}
\usepackage{titling}
\usepackage[caption2]{ccaption}
\usepackage{xr-hyper}
\usepackage{hyperref}
\usepackage{nameref}
\usepackage{zref-xr}
\zxrsetup{toltxlabel}

\usepackage[utf8]{inputenc}
\usepackage{pdflscape}
\usepackage{amssymb,amsbsy,color,amsmath}
\usepackage{graphicx}
\usepackage{dcolumn}
\usepackage{pgfplotstable}
\usepackage{filecontents}
\usepackage{longtable}
\usepackage{authblk}
\usepackage{hyperref}

\usepackage{subcaption}

\usepackage{authblk}
\usepackage{pdfpages}
\pdfoutput=1
\begin{document}
%
\title{Morphology of travel routes and the organization of cities}
\newcommand{\RochesterP}{Department of Physics and Astronomy, University of Rochester, Rochester, NY, USA}
\newcommand{\RochesterC}{Goergen Institute for Data Science, University of Rochester, Rochester, NY, USA}
\newcommand{\Korea}{Department of Energy Science, Sungkyunkwan University, Suwon, South Korea}
\newcommand{\Tokyo}{Institute of Innovative Research, Tokyo Institute of Technology, Tokyo, Japan}
\newcommand{\MIT}{MIT Media Lab, Massachusetts Institute of Technology, Cambridge, MA, USA}
\newcommand{\SFI}{Santa Fe Institute, Santa Fe, NM, USA}

\author[1]{Minjin Lee}
\author[2]{Hugo Barbosa}
\author[3,4]{Hyejin Youn}
\author[5]{Petter Holme}
\author[2,6]{Gourab Ghoshal}

\affil[1]{\Korea}
\affil[2]{\RochesterP}
\affil[3]{\MIT}
\affil[4]{\SFI}
\affil[5]{\Tokyo}
\affil[6]{\RochesterC}


\date{}
\makeatletter
\newcommand*{\toccontents}{\@starttoc{toc}}
\makeatother

\maketitle
\toccontents


	

\newpage
\phantomsection
\addcontentsline{toc}{chapter}{Manuscript}

%
%
%




%

\maketitle
\begin{abstract}
The city is a complex system that evolves through its inherent social and economic interactions. 
Mediating the movements of people and resources, urban street networks offer a spatial footprint of these activities; consequently their structural characteristics have been of great interest in the literature. In comparison, relatively limited attention has been devoted to the interplay between street structure and its functional usage, i.e., the movement patterns of people and resources. To address this, we study the shape of 472,040 spatiotemporally optimized travel routes in the 92 most populated cities in the world.  The routes are sampled in a geographically unbiased way such that their properties can be mapped on to each city, with their summary statistics capturing mesoscale connectivity patterns representing the complete space of possible movement in cities.
The collective morphology of routes exhibits a directional bias that could be described as influenced by the attractive (or repulsive) \emph{forces} resulting from congestion, accessibility and travel demand that relate to various socioeconomic factors. To capture this feature, we propose a simple metric, \emph{inness}, that maps this force field.  An analysis of the morphological patterns of individual cities reveals structural and socioeconomic commonalities among cities with similar inness patterns, in particular that they cluster into groups that are correlated with their size and putative stage of urban development as measured by a series of socioeconomic and infrastructural indicators. Our results lend weight to the insight that levels of urban socioeconomic development are intrinsically tied to increasing physical connectivity and diversity of road hierarchies. 
\end{abstract}


\noindent {\large{\bf Introduction}}
\vspace{5px}

 The city is an archetype of a complex system, existing, and evolving due to the myriad socioeconomic activities of its inhabitants~\cite{Bettencourt2010, Pan:1fh, Batty:2012dw}. These activities are mediated by the accessibility of urban spaces depending on the city topography and its infrastructural networks~\cite{Sim20150315}, a key component of which are the roads. Indeed, different street structures result in  varying levels of efficiency, accessibility, and usage of the transportation infrastructure~\cite{youn2008, Cardillo:2006fk, Justen:2013kl, Witlox:2007eq, daFCosta:2010er,  Wang:2012jn, Kang:2012by}. Structural characteristics, therefore, have been of great interest in the literature~\cite{Wang:2011ds,Rui:2013jd, Louf:2014jz, Strano:2013hf} and many variants of structural quantities have been proposed and measured 
in urban contexts, including the degrees of street junctions~\cite{Masucci:2009ja}, lengths of road segments~\cite{Strano:2013hf}, cell areas or shapes delineated by streets~\cite{Rui:2013jd}, anisotropies~\cite{Louf:2014jz}, and network centrality~\cite{Jiang:2007ch, Crucitti:2006haa}. Collectively, these structural properties have uncovered unique characteristics of individual cities as well as demonstrated surprising statistical commonalities manifested as scale invariant patterns across different urban contexts~\cite{Barthelemy:2011dq, Batty:2007:CCU:1543541, Goh:2016fg}.

While these studies have shed light on the statistical structure of street networks, there is limited understanding of the interplay between the road structure and its influence on the movement of people and the corresponding flow of socioeconomic activity; that is, the connection between urban dynamics and its associated infrastructure~\cite{Sun20141089}. One way to tease out this connection is to examine the \emph{sampling of routes}, that is an examination of how inhabitants of a city potentially utilize the street infrastructure. A number of studies have conducted research on the empirical factors behind the choice of routes~\cite{Lima:2016ima, Thomas:2015fv, DeBaets:2014em, Viana:2013cq, Patterson_2016}, yet comparatively little effort has been devoted to their geometric properties, that is their \emph{morphology}.

Indeed, the morphology of a route is shaped by the embedded spatial pattern of a city (land use and street topology) in association with dynamical factors such as congestion, accessibility and travel demand which relate to various attendant socioeconomic factors. Analyzing the morphology of routes, therefore, allows us to \emph{potentially} uncover the complex interactions that are hidden within the coarse-grained spatial pattern of a city. Furthermore, the morphology also encodes the collective property of routes, including their long-range \emph{functional} effects. For example, a single street, depending on its connectivity and location, can have influence that spans the dynamics across the whole city (Broadway or Fifth avenue in NYC for instance).

In particular, traffic patterns and the shape of routes have been shown to be determined, among other factors, by two competing forces~\cite{Thomas:2015fv}. On the one hand, one finds an increased tendency of agglomeration of businesses, entertainments and residential concerns near the urban center, correspondingly leading to a higher density of streets~\cite{Barthelemy:2008ft, Clark:2016vj} and thus attracting traffic and flows \emph{towards} the interior of the city (positive urban externality). Conversely, this increasing density leads to congestion and increase in travel times (negative urban externality) thus necessitating the need for arterial roads or bypasses along the urban periphery to disperse the congestion at the core. This has the effect of acting as an opposing force, diverting the flow of traffic \emph{away} from the interior of the city.

\begin{figure}[htp]
	\centering
\includegraphics[width=0.7\textwidth]{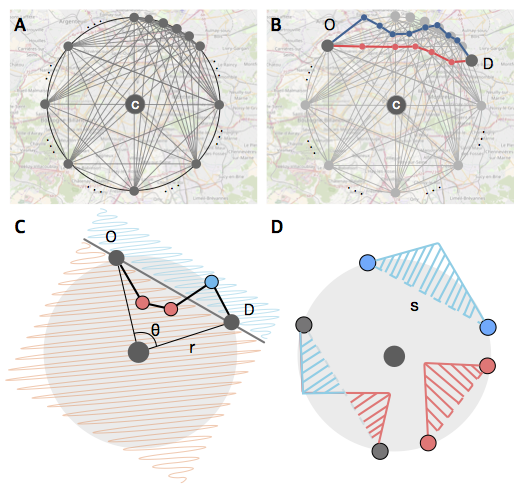}
\caption{{\bf Data sampling and definition of Inness $I$}
{\bf A} 36 origin-destination (OD) pairs (spaced out at intervals of $10^{\circ}$) are assigned along the circumference of circles at a distance of 2km, 5km, 10km, 15km, 20km and 30km from the city center C. {\bf B} For each OD pair, we query the OpenStreetMap API and collect the shortest routes (red) and the fastest routes (blue) (shown here for a representative OD pair in Paris). {\bf C} A typical OD pair with the straight line connecting them representing the geodesic distance $s$; $r$ is the radial distance from the center and $\theta$ is the angular separation relative to the center. We define the Inness ($I$) to be the difference between the \emph{inner} travel area (polygon delineated by red inner point and straight line) and the \emph{outer} travel area (polygon delineated by blue outer point and straight line). {\bf D} Three possible route configurations between multiple OD pairs. One with an exclusively outer travel area (blue), one with an exclusively inner travel area (red) and one where there is some combination of both.}
\label{fig:data_sampling}
\end{figure}

We investigate these competing effects through a detailed empirical study of the \emph{shape} of 472,040 travel routes between origin-destination points in the 92 most populated cities in the globe, representing all six inhabited continents. Each route consists of a series of connected roads, accompanying information on their geographical location, length, and speed limit retrieved from the OpenStreetMap database~\cite{OpenStreetMap}. We split our analysis between the shortest routes (necessarily constrained by design limitations and city topography) and the fastest routes (representing the effects of traffic and dynamic route sampling), with the former representing aspects of the city morphology, while the latter in some sense representing the dynamics mediated by the morphology (see Methods and Fig.~\ref{fig:data_sampling} for details). Specifically, the shortest routes are a function of the \emph{bare} road geographic structure, while the fastest routes represent the \emph{effective} geographic structure---a function of the heterogeneous distribution of traffic velocity resulting from varying transportation efficiency and congestion patterns \cite{Morris:2012bs, Strano:2015hr, Ashton:2005ck, Jarrett:2006ha}. 
 
To uncover the functional morphology of these two categories of routes, we define a novel geometric metric, which we term \emph{inness}---a function of both the direction and spatial length of routes---that captures the tendency of travel routes to gravitate towards or away from the city center. This metric serves as a proxy for the geographical distribution of \emph{hidden} attractive forces that may be implicit in the sampling of streets (as reflected in directional bias) and that otherwise cannot be captured by existing measures. Our analysis represents a step towards the very important challenge of determining the spatial distribution of urban land-use and street topology to balance the inherent negative and positive urban externalities that result from rapid urbanization~\cite{Thomas:2015fv}.

\section*{Results}
\label{sec:results}


\subsection*{Definition of Inness $I$}
\begin{figure}[htp]
	\centering
\includegraphics[width=0.9\textwidth]{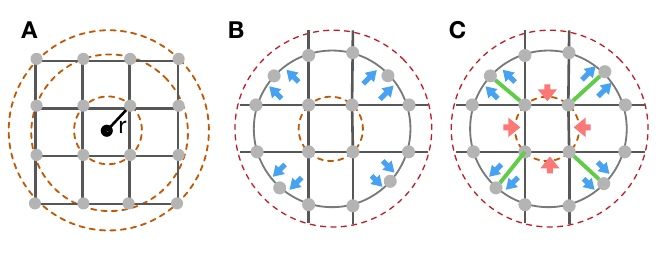}
\caption{{\bf Biasing forces found in urban morphology} Three schematic urban street arrangements share similar topological structure, but different geometric layouts resulting in varying dynamics.
 {\bf A} A grid structure where shortest paths between points at the same radius shows no directional bias. {\bf B} Repulsive forces relative to the origin (marked in blue) emerge as we break the grid symmetry by relocating the four outer points on the inner equidistant ring-line. Paths lying on this ring now have shortest paths that traverse the periphery and avoid the center. {\bf C} Further perturbing the topology by increasing connectivity to the center (marked as four green lines) now leads to shortest paths that go through the center as if an attractive force is present (marked in red).
}
\label{fig:forces}
\end{figure}

Fig.~\ref{fig:forces} illustrates schematics of the \emph{forces} hidden in the city's morphological patterns, shaping, and shaped by infrastructure and socioeconomic layouts. For the case of a square grid, as shown in Fig.~\ref{fig:forces}A, shortest routes between any two points at a distance $r$ either correspond trivially to the line connecting them directly, or are degenerate paths that traverse the grid in either direction. Taking the average of the multiple paths cancels any directional bias relative to the center of the grid. Yet, a small perturbation of this regularity can change this neutral feature dramatically, as shown in Fig.~\ref{fig:forces}B, where we shift the four outermost points inwards as to place them on the second ring from the center.  Points lying on this ring have shortest routes that lie along the periphery, thus introducing a dispersive force away from the center (marked as blue arrows). In Fig.~\ref{fig:forces}C, we further perturb the topology by adding four lines from the outer ring to the inner ring (marked in green) thus increasing connectivity towards the center. Shortest paths between pairs on the outer ring traverse through the inner ring and are curved towards the city center, resulting in an attractive force (represented as red arrows). Beyond this simple example, which is primarily a function of topology and is applicable to shortest paths, other factors are in play such as travel time and route velocity as considered in ~\cite{Thomas:2015fv}, that will necessarily affect the patterns seen in fastest paths. Furthermore, the illustration assumes a single center of gravity, as it were, whereas such effects may manifest itself at multiple scales, resulting in cancellation of any measurable force towards a putative city center.

To capture whether such an effect manifests itself at the scale of the city, or is indeed neutral due to the ``detuning" at smaller scales, we define a metric called the \emph{Inness} $I$. Fig.~\ref{fig:data_sampling}C illustrates how a typical route between any OD pair can be divided up into segments that are directionally biased towards, or away from the city center as measured relative to the geodesic distance $s$ between the pair. We label points lying closer to the center than the geodesic {\em inner points}, while those lying further away are {\em outer points}. For example in the schematic shown in Fig.~\ref{fig:data_sampling}C points located in the pink shaded area are inner points, and those on the opposite side (shaded blue) are outer points. We define an \emph{inner travel area} delineated 
by the polygon of inner points and the geodesic line, to which we assign a positive sign. Conversely, an \emph{outer travel area} is defined by the geodesic line and the collection of outer points, whose sign is negative. Having adopted this convention, $I$ is the difference between the inner area and outer travel areas, 
\begin{equation}
I = A_{\textrm{in}} - A_{\textrm{out}}, 
\label{eq:innesss}
\end{equation}
which can be calculated using the shoelace formula for polygons (see Methods). In Fig~\ref{fig:data_sampling}D we show three possible idealizations of a route; one with only outer travel area (blue), one with only inner travel area (red), and one with a mixture of both outer and inner travel areas (combination of blue and red). 

The inness of a node in the road network is a result of aggregating characteristics of all possible routes that pass through that point. Indeed, it reflects the structure of the network since it is a metric influenced by topology and network connectivity.  However, in addition to this, it captures geometric aspects of the network since it is a measure based on the curvature of the roads along a route and encodes structural information at both the global and local scales. In that sense, one can consider it to contain elements of various standard structural network metrics (see Sec.~\ref{si:sec:network} for details and comparison with a series of network metrics).

\subsection*{Average Inness for shortest and fastest routes}

\begin{figure*}[htp]
	\centering
\includegraphics[width=0.7\linewidth]{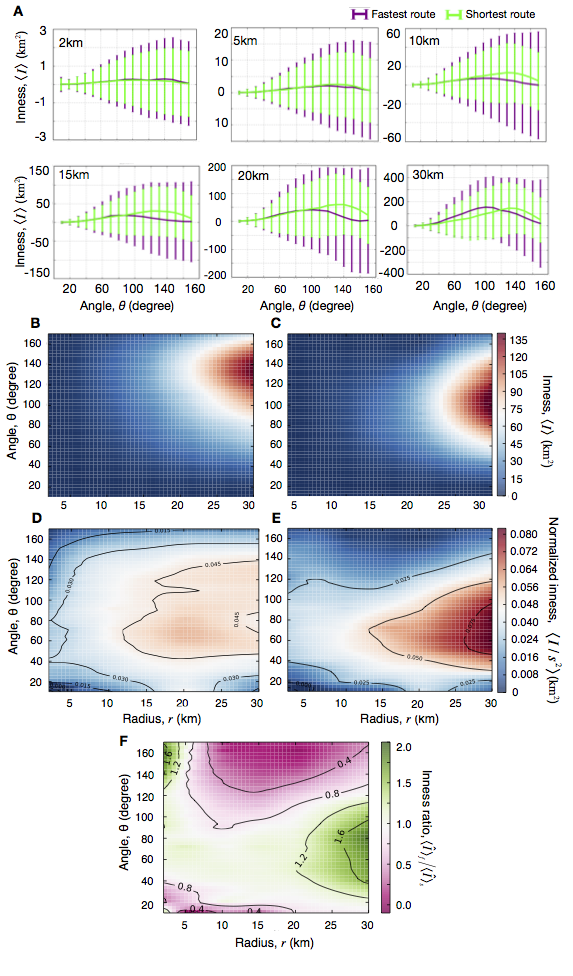}
\caption{{\bf The average inness across 92 cities}: {\bf  A} The average and standard deviation of $I$ as a function of $\theta$ shown for multiple radii $r$ measured from the city center. The curve for the shortest route is shown in green and the fastest route in purple. The density plot of $\langle I \rangle$ in function of $r$ and $\theta$ for the shortest routes {\bf B} and fastest routes {\bf C}. The normalized or dimensionless inness $\langle \hat {I} \rangle  = \langle I/s^2 \rangle $ for the shortest {\bf D} and fastest routes {\bf E}. {\bf F} Ratio of the normalized inness of fastest  $\langle \hat I \rangle_f$ and shortest routes  $\langle \hat I \rangle_s$}.
\label{fig:avg_inness}
\end{figure*}

We begin our analysis by an examination of the qualitative trends of $I$. In Fig.~\ref{fig:avg_inness}A, we plot the average inness $\langle I \rangle$ (averaged over the 92 cities) for both the shortest (green curves) and the fastest routes (purple curve) as a function of the angular separation $\theta$, for multiple radii $r$. In the vicinity of the city center (2-5km), we see a neutral trend for the shortest routes ($\langle I \rangle \approx 0$) although fluctuations increase with angular separation ($80^{\circ} \leq \theta \leq 160^{\circ}$). 
The fluctuations anticipate a clear \emph{inward bias} that emerges at a distance of $r \geq 10$ km, visible as a pronounced positive peak in $\langle I \rangle $ that grows progressively sharp with increasing $r$. The qualitative trend of $\langle I \rangle$ is indicative of the presence of a core-periphery street network structure present in varying degrees across all cities \cite{Lee:2014jn} and suggests that the \emph{attractive forces} introduced in Fig.~\ref{fig:forces} tend to manifest themselves at the scale of the entire city, pointing to the existence of an \emph{effective center} (on average). Indeed, for fixed $r$, the geodesic distance $s$ between any OD pair is a monotonically increasing function in $\theta$.  The longer the distance, the more likely it is for the route to drift towards the center---due to greater connectivity in the center compared to the periphery. This is a possible explanation for the observed inward bias and is indicative of a progressively lower density of streets in the periphery. 

The roads in our dataset are organized in a hierarchical fashion consisting of motorways, primary and trunk roads at the top of the hierarchy and residential and service roads being at the lower levels (Fig. \ref{si:fig:road_types_distributions}). The shortest paths considered thus far are primarily composed of secondary and residential roads~(Fig. \ref{si:fig:road_types_shortest});  conversely the fastest routes tend to use a smaller subset of the overall network, primarily motorways that are in general major highways with physical divisions separating flows in opposite directions such as freeways and Autobahns (Fig.~\ref{si:fig:road_types_fastest}). This heterogeneity in the road capacity is bound to introduce differences in the inness profiles of shortest and fastest routes. Reflecting this, one sees an inward bias for the fast routes emerging around 10km, but markedly less pronounced than seen for shortest routes, although the qualitative trend of increasing inward bias with $r$ is maintained. The lower inward bias of faster routes can be explained by the fact that the motorways are typically located in the periphery of cities.  Additionally, the angular range of the observed inward bias is \emph{lower} than that seen for shortest paths ($45^{\circ} \leq\theta\leq 120^{\circ}$) and the fluctuations are significantly larger. This is  indicative of the heterogenous spatial distribution of velocity profiles in the primary roads across cities (due to varying levels of infrastructure), coupled with the fact that they are in general longer than secondary roads.

The collective angular and radial dependence of $\langle I \rangle$ are shown as density plots for the shortest and fastest routes in Figs~\ref{fig:avg_inness}B\&C. The monotonic increase of $\langle I \rangle $ with $r$ is apparent in both cases particularly at $r \sim 15$km (also shown explicitly in Fig.~\ref{si:fig:summary_inness_stats}A). The differences in angular dependence can be clearly seen with the fastest routes having a sharp $\langle I\rangle$ at a lower angular range than shortest paths. Notable is the absence of any outward bias (negative values of $\langle I \rangle$) at any radial or angular range. 

\subsection*{Dimensionless Inness $\hat I$}
The observed trends for the inness do not take into account the effects of varying travel areas at different distances from the center, or indeed the variation in urban size across the studied cities. To account for these effects we note that the travel area (averaged across cities) increases roughly quadratically with the geodesic distance $s$ (Fig.~\ref{si:fig:summary_inness_stats}K), a trend also observed in~\cite{Lima:2016ima} where the characteristic shapes of city routes had travel areas of $\textrm{O}(s^2)$. 
Therefore, to account for  any bias from variations in travel area within and across cities we define a rescaled inness
\begin{equation}
\hat {I} = \frac{I}{s^2}.
\label{eq:norm_inness}
\end{equation}
 In Figs.~\ref{fig:avg_inness}D\&E, we plot $\langle \hat I \rangle$ for the shortest and fastest routes finding that the inness effect is robust to potential biases due to length or area of trips. While the qualitative behavior is similar to that seen for $\langle I \rangle$, the inward tendency of routes is present over a \emph{broader} $r$ and $\theta$ for both the shortest and fastest routes. For instance, inward biases are apparent at distances of 5km from the city center, an effect suppressed in $\langle I \rangle $ due to the correspondingly smaller travel area. Furthermore, we now find a relatively more homogeneous distribution with a comparatively weaker dependence on $r$ and $\theta$. The trend for $\langle \hat I \rangle $ continues to support an average core-periphery structure in the cities we study (and therefore a city center), in combination with a smoothly decreasing density distribution of streets away from this center. The weaker angular dependence, in particular, hints at an isotropic variation in density of street junctions.
 
The distribution of $\langle \hat I \rangle $ is comparatively less homogeneous for fastest routes, primarily due to the predominance of the high-capacity roads  (Fig. \ref{si:fig:road_types_distributions}) adding, therefore, more variation to the inness profile.  Across a wide range of $r$ and $\theta$, $\langle \hat I \rangle$ is generally lower than shortest routes, while there is sharp increase at $15\leq r\leq30$ and $40\leq\theta\leq100$. This is likely due to the some of the motorways being specialized structures such  as \emph{ring roads} or \emph{bypasses}, that serve as attractors for traffic in the city periphery. This is confirmed by plotting the ratio $\langle \hat I \rangle_f /\langle \hat I \rangle_s$ in Fig.~\ref{fig:avg_inness}F, where one sees a factor of two or more inward bias in the fastest routes as compared to the shortest routes near the city periphery ($\sim25$km). 




\subsection*{Inness distribution for individual cities}

Having examined the properties of average directional biases across urban areas, we now turn our attention to the patterns in individual cities. Indeed, as the fluctuations in Fig.~\ref{fig:avg_inness}A show, there is variability in the inness pattern across cities, reflecting the differences in the level of road hierarchies and organization. The composition of the shortest and fastest routes in terms of different road hierarchies varies significantly from city to city. For example, in some cities (Atlanta, Houston, Madrid) the fastest routes tend to be through motorways, whereas in others (Luanda, Kolkata and Pune) they tend to be composed of \emph{primary} roads (Figs.~\ref{si:fig:road_types_fastest}); these differences are likely to affect their respective inness profiles.

\begin{figure*}[htp]
	\centering
\includegraphics[width=\linewidth]{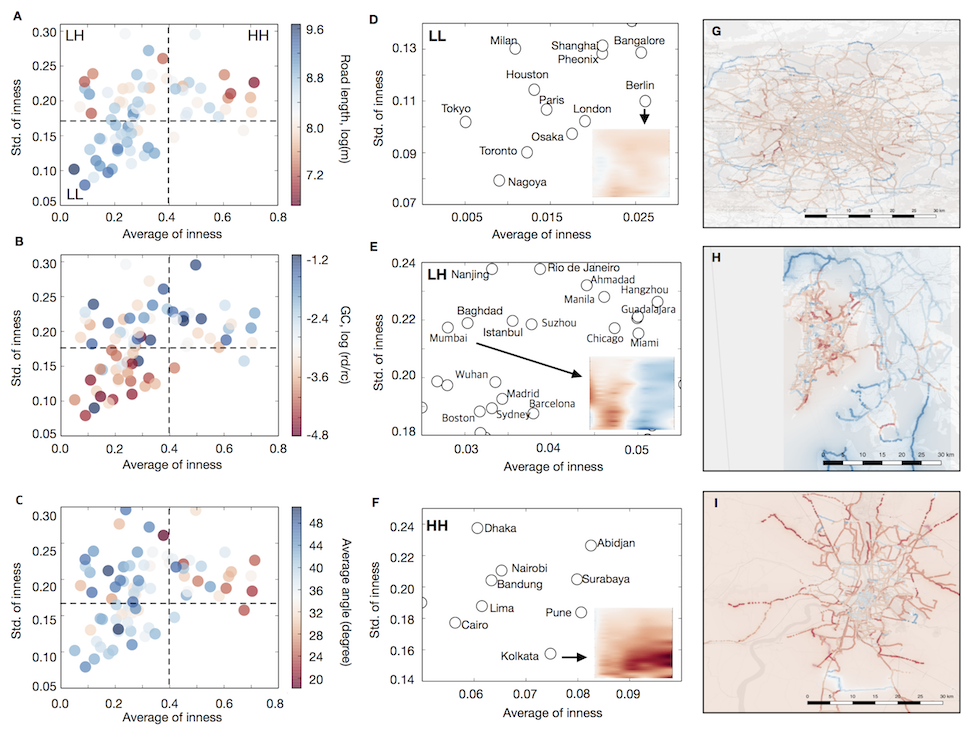}
\caption{{\bf The average and standard deviation and spatial distribution of the shortest path inness for individual cities.} The standard deviation plotted in function of average inness for each city. The cities are divided into three groups by their value of average and standard deviation of inness; Low-Low(LL), Low-High(LH), High-High(HH). The color of points indicate the road length {\bf A}, the level of geographical constraint (GC) {\bf B}, and a measure of peripheral connectivity {\bf C}. We enlarge three zones marked LL, LH and HH and label the cities explicitly in panels {\bf D}, {\bf E}, {\bf F} as well as display in inset $\hat I$, for representative cities in each region (Berlin, Mumbai, Kolkata). In {\bf G}, {\bf H} and {\bf I} we plot the spatial distribution of $\hat I$ projected on to the physical maps for the three representative cities. The color of street intersections correspond to the average $\hat I$ of all routes passing through the intersection,  with values in the interval $-0.3 \leq \hat I \leq 0.3$ and ranges from blue to red with increasing $\hat I$.}

\label{fig:avg_stdev_inness_dist}
\end{figure*}

To investigate the effect of these differences, we plot each city as a function of the standard deviation and the average of $\hat I$ for the shortest paths (Fig.~\ref{fig:avg_stdev_inness_dist}A--C). A cluster of cities appear in the range $0.0 \leq \langle \hat I \rangle \leq 0.08$ with some outliers at both positive and negative values. Ignoring the outliers for the moment, roughly speaking, we identify three regions; low average and low standard deviation (LL), low average and high standard deviation (LH) and high average and high standard deviation (HH). (See Figs.~\ref{si:fig:som_shortest}\&\ref{si:fig:som_fastest} and Sec.~\ref{si:sec:outliers} for details on individual and outlier cities.)  Each city is also colored according to three metrics reflecting infrastructural and geographical  features: In~\ref{fig:avg_stdev_inness_dist}A, we show the total length of  roads within 30km from the city center; in~\ref{fig:avg_stdev_inness_dist}B we show a measure of geographical constraints (GC) that captures the presence and size of barriers such as rivers, coastlines, mountains or  industrial facilities;  finally in~\ref{fig:avg_stdev_inness_dist}C we show a measure of peripheral connectivity acting as a proxy for the presence of ring roads in the city (details for each metric in Sec. \ref{si:sec:roads_metrics}). In Fig.~\ref{fig:avg_stdev_inness_dist}D--F we show a selection cities from each region, along with the density plot of $\hat I$ from a representative city shown as inset.

It appears that cities within each region tend to share some common features with respect to these metrics and their inness profiles. Those in the LL group tend to have longer total length of roads, fewer geographic constraints and strong peripheral connectivity, indicating high levels of infrastructural development. The inness profile is typically neutral (as can be seen for Berlin), indicating no discernible center of the city to which routes are drawn. On the other hand cities in the HH group tend to have shorter road lengths, limited connectivity in the periphery (indicating relatively lower levels of infrastructural development) and more geographical constraints than the LL group. These cities also have a markedly positive inness profile (shown for Kolkata) suggesting that navigability of the city passes through a central core. The LH group appears to show a combination of high and low values in terms of the infrastructural metrics, yet is notable in showing markedly higher geographical constraints than cities in the other regions. This is reflected in a rather peculiar inness profile which manifests itself positively at short distances, yet is negatively at longer ranges.

To investigate these trends further, we plot the spatial distribution of $\hat I$ on a geographic map of a representative city from each region. The spatial distribution is generated by considering every intermediate point in a city route, and calculating the average $\hat I$ of all routes that pass through this particular location. In Fig~\ref{fig:avg_stdev_inness_dist}G, we show the spatial distribution of $\hat I$ for Berlin. Throughout a wide swathe of Berlin, we find a homogeneous distribution of moderately positive (almost neutral) inness (red), with roads near the city boundary showing a marginally negative inness value (blue). Similar patterns are observed in other large urban agglomerations such as Tokyo and Paris (Figs.~\ref{si:fig:spatial_distribution_of_cities}A\&C). By and large these cities are large urban areas with advanced infrastructure and strong levels of connectivity.

Next, we focus on the LH group of cities, those with a mix of inward \emph{and} outward bias in the route patterns. In Fig~\ref{fig:avg_stdev_inness_dist}H, we show the spatial profile for Mumbai which displays two distinct regions with inner and outer bias. The two regions are separated by the Arabian Sea, and connected by (a few) bridges. The left hand side of the map corresponds to the more densely connected part of Mumbai (its economic center) thus most routes within this region have an inward bias. The appearance of the outward bias in the other region is due to the lack of direct connectivity with the economic center, as routes have to pass through one of few bridges that connect the island to the mainland and are thus subject to considerable detour. A similar pattern is seen in other cities in this group, almost all of which have geographic barriers (rivers, seas, hills, mountains) that either spread out across the city, or divide the city into distinct regions. Additionally there are cities in this group  with no geographical barriers, but \emph{artificial} barriers attributing a similar effect on route profiles. A notable example of this is Miami which has a prominent rock-mining industrial site in the Western Miami-Dade County (See Figs.~\ref{si:fig:spatial_distribution_of_cities}E\&G for this and other examples).

Finally, we examine the spatial distribution of $\hat I$ for Kolkata as a city in the HH category, shown in Fig~\ref{fig:avg_stdev_inness_dist}I.
It is apparent that the profile is more distinct than that seen for Berlin. There is a \emph{hub and spoke} type pattern with the spokes showing high levels of inness, presumably due to its function connecting outer regions to the city center. Indeed, a clear city center is apparent with limited to no connections across the periphery.  A similar pattern is seen for other cities in this group (Cairo, Medan, Figs.~\ref{si:fig:spatial_distribution_of_cities}I\&K).  There are two possible causes for this: either these cities have a relatively smaller \emph{effective} urban area (regions of strong connectivity), or they are large urban agglomerations with limited or underdeveloped infrastructure.


\subsection*{Differences between shortest and fastest routes in cities}

\begin{figure*}[htp]
	\centering
\includegraphics[width=\linewidth]{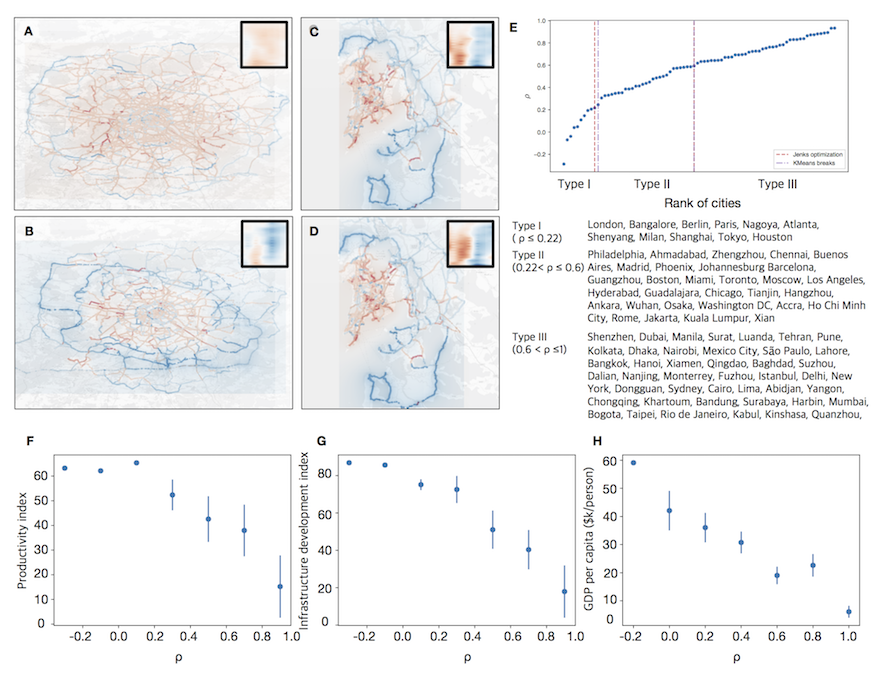}
\caption{{\bf Difference between shortest routes and fastest routes} Shortest routes {\bf A} and fastest routes {\bf B} for Berlin. Shortest routes {\bf C} and fastest routes {\bf D} for Mumbai. The insets show the density plots of $\hat I$ with the same range as in Fig.~\ref{fig:avg_stdev_inness_dist}.
{\bf E} Pearson correlation coefficient $\rho$ between the inness patterns of shortest and fastest routes for each city. Cities are categorized into three groups (marked by vertical dashed lines) based on a K-means clustering and Jenks natural breaks optimization, conditioned on their level of correlation $\rho$. Three socio-economic indexes; Productivity index {\bf F}, Infrastructure development index {\bf G}, and GDP per capita {\bf H} plotted as a function of $\rho$ showing a clear monotonically decreasing trend. Points are averages over cities binned in intervals of 0.2, and bars represent the standard error. }
\label{fig:corr_short_fast}
\end{figure*}

The measured connection of inness profiles with infrastructural indicators suggests that more information can be gleaned by studying the differences between shortest and fastest routes. Indeed, while the former is more connected to spatial and geographical constraints, the latter shares a more natural connection with developmental indicators. 
To better quantify this difference, we measure the Pearson correlation coefficient $\rho$ between $\langle \hat I \rangle_f$ and $\langle \hat I \rangle_s$, for each city, shown in Fig.~\ref{fig:corr_short_fast}E in increasing order from negative to positive values. Shown as dashed vertical lines are the result of using $K$-means clustering and Jenks natural breaks optimization to partition the cities into three groups with both methods producing nearly identical divisions. (Alternative clustering approaches revealed similar results, see Sec.~\ref{si:sec:clustering}.)

To investigate this partitioning, we pick a representative city from each end of the spectrum: Berlin with $\rho < 0$ and Mumbai with $\rho \approx 1$. 
Figs.~\ref{fig:corr_short_fast}A\&B show the spatial distribution of $\hat I$ in Berlin for the shortest and fastest routes. 
Unlike the neutral inness seen for shortest routes, fastest routes display a strong outward bias starting from a radial distance of 15km, which seems to be a consequence of ring-like arterial roads dispersing traffic away from the center. Indeed, this supports our metaphor of competing forces sketched in Fig.~\ref{fig:forces}; the high connectivity of streets in the center of Berlin tends to draw flow towards the city (witnessed by the mildly positive inness profile of the shortest routes)  while the faster roads pushes routes outwards. A similar trend is seen for all cities with a negative correlation (Tokyo and Paris shown in Fig.~\ref{si:fig:spatial_distribution_of_cities}A-D), with the presence of arterial ring roads (built presumably to alleviate congestion) near the city periphery being the main driver of the differences (in line with higher values of peripheral connectivity measured earlier). A majority of these cities, that are members of partition Type I, correspond to the those seen in the LL group in Fig.~\ref{fig:avg_stdev_inness_dist}B.
 
Unlike Berlin, Mumbai exhibits virtually identical profiles of inness between shortest and fastest routes, as seen in Figs.~\ref{fig:corr_short_fast}C\&D,
with fewer arterial roads or bypasses that can divert traffic away from the city center. In the case of Mumbai, this is due to salient geographic constraints, but a similar pattern is also seen in other cities like Kolkata, that also lack peripheral roads. Thus, cities that either suffer from some kind of geographic constraint, or relatively underdeveloped infrastructure tend to show a higher correlation between the two types of routes. These cities in partition Type III, include the majority of cities in the HH group and a few from LH groups in Fig.~\ref{fig:avg_stdev_inness_dist}.

Cities with intermediate correlation in partition Type II, tend to be those with a profile seen in the LH group in Fig.~\ref{fig:avg_stdev_inness_dist}. The behavior seen here seems to be some combination of what drives the trends seen in Type I and Type III cities. As it happens, however, there is the exceptional case of New York. The city shares the same features as Type I cities, \emph{i.e.}\ it is a large urban area, with highly developed infrastructure, yet, there appears to be a strong correlation between the fastest and shortest routes (Fig.~\ref{si:fig:new_york}). This is likely due to the unique geography of motorways in the New York metropolitan urban area, which unlike typical Type I cities, does not have ring-like motorways in the periphery. Instead New York consists of a series of radial and grid-like motorways whose overall effect is to cancel any observable directional bias.

Advanced levels of infrastructure are typically reflected in improvements across a variety of socioeconomic indices. To examine whether the measured behavior of the inness captures any of this, we consider three socio-economic indicators: the productivity index sourced from the City Prosperity index (CPI) created by the UN~\cite{CPI}, the infrastructure development index (also sourced from CPI) and finally the GDP-per-capita ~\cite{GDP}. In Figs.~\ref{fig:corr_short_fast}F-H we plot these metrics as a function of the correlation coefficient $\rho$ that was used to cluster the cities. Remarkably in all three cases, there is a clean monotonic decrease of the indicators with increasing $\rho$, suggesting that inness also encodes information on socioeconomic development. A relatively clear pattern emerges whereby the majority of Type I cities are large urban agglomerations with advanced infrastructure and strong socioeconomic indicators, Type III cities by and large have comparatively limited infrastructural and socioeconomic development, and finally Type II cities share a combination of these features.

%

\section*{Discussion}

Networks of streets and roads are the primary facilitators of movement in urban systems, allowing residents to navigate the different functional components of a city. Since navigability is a key ingredient of socioeconomic activity, street networks represent one of the key (if not the most important) infrastructural components. In particular the utilization of street networks captures  the complex interactions between people, and the flow of goods and services in urban systems. However, there is relatively limited understanding of this facet as existing macroscopic or microscopic measures are not able to fully capture its properties and associated effects.  Part of the challenge is the limited availability of detailed and high-resolution data of dynamics taking place on such networks, necessitating a choice for investigative studies to be made in terms of granularity or scale. In this manuscript we erred on the side of the latter, and conducted a systematic mesoscale analysis of street morphology---representing a proxy for the potential dynamics---through the introduction of a novel metric that we term inness. The inness encapsulates the direction, orientation and length of routes, thus revealing the morphology of connectivity in street networks, including the implicit infrastructural and socioeconomic forces that may inform routing choices. 

The average inness pattern points towards the existence of a core-periphery structure across the majority of cities with a high density of streets in the city center with a progressively lower density as one moves towards the periphery.  This pattern is particularly seen among Type III cities mostly corresponding to those in Figs.~\ref{fig:avg_stdev_inness_dist}D and \ref{si:fig:spatial_distribution_of_cities}I--L. These happen to be the most numerous in our sample, and therefore dominate the average statistics. Their spatial distribution is characterized by a hub-spoke structure, with Kolkata being an archetype, and have a strong correlation between the patterns of fastest and shortest routes (Fig.~\ref{fig:corr_short_fast}I). Given that many (but not all) cities in this category are in developing countries (as confirmed by GDP-per-capita and prosperity indices), this feature seems to indicate a relatively underdeveloped infrastructure (confirmed by lower values of infrastructural development index) with the absence of bypasses, highways, or ring-roads that disperses traffic more efficiently.

Interestingly these qualitative features capture \emph{certain elements} of the classical hypothesis of \emph{central place theory} advanced by Christaller~\cite{christaller1966central}. The theory postulates that cities are organized into a hierarchy of cores that perform specific economic functions depending on their position in the hierarchy~\cite{10.2307/142299}. As the core at the highest level of the hierarchy (which is usually located near the city center) has the most diverse and complex economic functions, lower cores that perform less complex functions need to connect to the higher core to meet the demand of some of the activities in the higher core. Furthermore cores at the lower levels of the hierarchy have minimal interactions between them.  

An alternative version advanced by L\"osch~\cite{losch1954economics} claims instead that cores at the same level develop specialized and symbiotic functions and thus develop interactions with each other, having the effect of dispersing activity from the center. Indeed, this feature seems to be present in our Type I cities  which compose the majority in Figs.~\ref{fig:avg_stdev_inness_dist}B and \ref{si:fig:spatial_distribution_of_cities}A--D. These are predominantly large urban areas with high levels of infrastructural and socioeconomic development (Figs.~\ref{fig:corr_short_fast}F-H). Cities such as Berlin, Paris, and London, display a relatively neutral inness trend across locations, pointing towards a uniform density of streets across the city. In particular, a comparison of the spatial distribution of inness for the fastest and shortest routes reveals the presence of ring roads connecting the peripheral areas of the city, and dispersing traffic away from the city center (Fig.~\ref{fig:corr_short_fast}E).  Considering one of the purposes of building peripheral roads is facilitating material movement around the city, the observed discrepancy between the shortest and fastest routes might reveal that a developed urban area decouples material (resource) movement from human movement via the introduction of peripheral roads. This is in line with the assertion that cities tend to transform their social and economic functionality into dematerialized operations as they develop \cite{Youn2016, Pumain2006, Bettencourt2014}. 

A number of cities fall between the spectrum of these two levels of categorization. These are predominantly Type II cities the majority of which are in the LH group in Figs.~\ref{fig:avg_stdev_inness_dist}E and ~\ref{si:fig:spatial_distribution_of_cities}E--H. Most (but not all) have geographical or artificial constraints within the city (Mumbai and Rio de Janeiro being notable examples), leading to a mixture of dense and poor connectivity between different locations. Indeed, some have advanced infrastructure (Miami), while others less so (Medan) and they also differ in terms of the size of the urban areas. These lie somewhere in between the Christallerian and Loschian classification, although one can take the argument only so far, absent other detailed dynamical information on socioeconomic activity, land use pattern as well as historical data on their evolution.



One must note that inness assumes the existence of a unique city center. Cities that show a strong positive value of inness, may thus be considered monocentric in the sense that there is a measurable center through which the majority of routes pass; while those with neutral inness are polycentric, in that there is no such center. Centricity in this case, has to be interpreted in the morphological sense reflecting the hierarchical organization and spatial patterns of roads.  Cities, in general, however, exist in a continuum between mono and polycentricity, depending on how one defines these terms. Generally, centricity in cities has been measured on the basis of locations of high population density or spatial patterns of land use~\cite{PhysRevLett.111.198702, louail2014mobile}, with no dominant quantitative definition for what constitutes the type of centricity~\cite{roth2011structure}. Indeed the notion of whether cities are monocentric or not, depends on the specific feature being investigated. It is notable, however, that some Type I cities considered to be polycentric in terms of our definition, are also similarly classified based on completely different metrics such as types of employment and their density patterns~\cite{PhysRevLett.111.198702, louail2014mobile,roth2011structure}. Though one must be careful with the analogy, given the limited basis of comparison, inness may be also interpreted as a parsimonious, and easily measured metric of centricity in cities, given its simple definition and intuitive interpretation.

Movement patterns are  intimately related with the spatial hierarchical distribution of economic activities~\cite{Batty2006}. While the self-similarity and hierarchy between industries in a city has been studied in detail~\cite{Youn2016}, the connection between the flow of people and goods to its economic hierarchy is less understood. While our study does not directly address this issue, it appears that inness as a metric encodes information on infrastructural organization in cities as well as certain aspects of socioeconomic activity. Given that it can be easily studied in large scale, augmenting our analysis with the spatiotemporal distribution of socioeconomic activities opens up a promising direction towards the understanding of urban structures and their evolution.

\section*{Methods}

\subsection*{Sampling Routing Pairs}

For each of our 92 cities, a city center is defined by referencing the coordinates from \url{latlong.net}~\cite{latlong} and the travel routes are sampled according to a choice of origin-destination pairs (OD) relative to the center and measured in spherical coordinates (distance from center $r$, and angular separation relative to center $\theta$). To avoid any sample bias, and to systematically investigate the dependence of route morphology on distance from the urban center, we only consider OD pairs at a fixed radius $r$. 

Furthermore at each $r$ we section the circumference of the circle at intervals of $10^\circ$ for a total of 36 points (with the minimal angular separation chosen to avoid effects of noise). We then vary the radius over the range 2km, 5km, 10km, 20km and finally 30km (roughly corresponding to a city boundary) and enumerate over all OD pairs by connecting the 36 points at a given radius $r$ for a total of $5 \times \binom{36}{2} = 3150$ total OD pairs. 

Finally, we query the OpenStreetMap API for the suggested travel routes connecting each of the pairs. In fact, for a better characterization of the functional features of the systems (e.g., road capacities) and the role of their hierarchical organizations, we obtained two different kinds of \emph{routes} between all these pairs: the \emph{shortest}, based on lengths of road segments, and the \emph{fastest} that accounts for both the length as well as the travel time based on flow capacity of the roads (i.e., speed limits, number of lanes etc.). A visual representation of our methodology in segmenting the city is shown in Fig.~\ref{fig:data_sampling}A and typical examples of the shortest and fastest route for a given city is shown in Fig.~\ref{fig:data_sampling}B. (For more details on our data samples, see Sec.~\ref{si:sec:materials}, Tables~\ref{si:tab:data_stats}\&\ref{si:tab:sample_fractions}.)

\subsection*{Calculation of Inness}
 The inness $I$ is calculated by summing over the areas of the number of polygons in the route by using the shoelace formula thus,
 \begin{equation}
 I = \frac{1}{2} \sum_{i=1}^{m} \textrm{sgn}(i) \left| \sum_{j=1}^{n}\det \begin{pmatrix}
 x_{i,j} & x_{i,j+1} \\
 y_{i,j} & y_{i,j+1}\\
 \end{pmatrix}
 \right|. 
 \label{eq:sholeace}
 \end{equation}
Here $n$ is the number of vertices of the polygon, $m$ is the total number of polygons in the route, 
$(x_{i,j}, y_{i,j})$ corresponds to the coordinate of $j$'th vertex of polygon $i$, 
and $\textrm{sgn}(i)$ accounts for our adopted convention for inner and outer points. 

%



\newpage
\setcounter{figure}{0}    
\phantomsection
\setcounter{tocdepth}{2}
\addcontentsline{toc}{chapter}{Supporting Information}

%
%
%
	\title{Morphology of travel routes and the organization of cities\\[.1cm]Supplementary Information}
%
%
	\maketitle
	\renewcommand{\listfigurename}{Supplementary Figures}
	\renewcommand{\listtablename}{Supplementary Tables}
	\renewcommand{\thesection}{S\arabic{section}}
	\renewcommand{\thefigure}{S\arabic{figure}}
	\renewcommand{\thetable}{S\arabic{table}} 
%
	%
	%
	%
	%
	
	%
	%
	

	\section{Materials and Methods}
	\label{si:sec:materials}
	\subsection{Collecting data from API service}
	Origin-Destination (OD) pairs were  generated having the cities' centers as the reference points. In the absence of a worldwide definition of a city center, we used the coordinates collected and provided by the \lq{latlong.net}\rq service, for each of the 92 cities. Having such coordinates as reference center points, we systematically computed \emph{theoretical} OD pairs corresponding to combinations of discretized radial and angular ranges. 
	
	In many cases, the theoretical points have no access to the streets networks, being therefore approximated to their closest point within the networks. Such approximation is done automatically by OSM routing API. However, to avoid the large discrepancies between a requested (theoretical) point and those returned by the OSM API, we applied two post-processing filters. First, if the distance of a returned OD coordinate is off by more than 1km from the center, we exclude such routes from the data. Second, we also excluded those routes whose lengths are longer than $3s+1km$, where $s$ is the geodesic distance between the origin and destination points. 
	
	\subsection{Data description}
	
	For each of the 92 cities, the maximum total number of unique driving routes is 630. However, after filtering out those discrepant routes and OD pairs, for each of the radii values, the average number of valid routes we analyzed were 575.9 (2km), 532.8 (5km), 461.7 (10km), 391.0 (15km), 349.3 (20km) and 254.7 (30km). Table S1. also describes the detour index (DI) (i.e., the ratio $d/r$ between the travel distance $d$ and geodesic distance $r$) for fastest (F) and shortest (S). Avg. waypoint represents the average number of route points as returned by the OSM API for each city.

	\newcolumntype{C}{>{\centering\arraybackslash}m{2.5em}}
	\pgfplotstableset{
		begin table=\begin{longtable},
			end table=\end{longtable},
	}
	\begin{filecontents*}{data_merge_final_1.csv}
City,short_df,fast_df,via,2km,5km,10km,15km,20km,30km
Abidjan,1.576,1.659,188.88,584,589,571,386,328,177
Accra,1.497,1.595,158.81,587,611,421,266,216,169
Ahmadabad,1.392,1.491,233.76,625,612,610,481,539,378
Ankara,1.48,1.599,367.45,612,619,580,571,420,397
Atlanta,1.307,1.478,386.32,599,621,624,629,630,630
Baghdad,1.6,1.766,212.38,550,602,606,496,339,220
Bandung,1.573,1.657,380.07,619,580,536,418,211,309
Bangalore,1.338,1.506,354.03,616,623,627,628,630,628
Bangkok,1.505,1.634,275.5,476,608,597,606,606,485
Barcelona,1.504,1.68,384.15,597,342,266,195,203,182
Belo Horizonte,1.525,1.682,456.71,612,617,619,606,470,516
Berlin,1.297,1.47,405.61,629,626,622,622,619,626
Bogota,1.566,1.661,455.55,608,621,606,583,583,427
Boston,1.374,1.519,626.85,576,616,571,520,495,405
Buenos Aires,1.353,1.511,178.25,613,520,259,252,231,209
Cairo,1.5,1.614,241.77,607,600,601,597,558,481
Cape town,1.471,1.562,275.8,479,357,251,320,187,66
Chennai,1.409,1.566,163.64,613,624,270,228,210,167
Chicago,1.387,1.551,206.56,590,346,228,210,210,190
Chongqing,1.72,1.769,287.77,448,521,578,480,403,304
Dalian,1.486,1.535,129.62,615,533,210,133,100,65
Dar es salaam,1.488,1.54,169.06,609,595,590,516,217,141
Delhi,1.409,1.483,262.27,618,604,608,617,621,453
Dhaka,1.568,1.648,244.44,519,593,462,489,380,293
Dongguan,1.569,1.61,159.03,587,500,512,425,542,516
Dubai,1.629,1.656,154.63,570,443,394,240,212,182
Fuzhou,1.656,1.719,481.95,608,586,503,492,469,426
Guadalajara,1.431,1.623,277.36,613,630,623,469,522,219
Guangzhou,1.55,1.644,312.96,512,558,594,556,537,502
Hangzhou,1.473,1.574,167.2,613,616,587,544,569,416
Hanoi,1.453,1.556,218.4,622,606,612,570,571,439
Harbin,1.593,1.638,119.13,543,456,406,243,425,178
Ho Chi Minh City,1.438,1.534,257.03,594,628,623,580,579,502
Houston,1.298,1.458,308.5,595,627,626,630,628,626
Hyderabad,1.38,1.533,346.95,623,620,622,621,622,546
Istanbul,1.462,1.615,335.32,454,569,431,452,315,187
Jakarta,1.461,1.618,236.13,604,620,620,551,348,246
Johannesburg,1.394,1.575,226.01,589,623,608,619,592,567
Kabul,1.656,1.678,219.18,351,508,376,179,161,27
Khartoum,1.538,1.615,228.41,603,606,339,357,405,217
Kinshasa,1.579,1.648,213.51,600,604,586,465,209,86
Kolkata,1.396,1.523,322.11,630,608,617,447,336,145
Kuala Lumpur,1.549,1.648,337.27,564,595,575,533,433,393
Lagos,1.633,1.726,283.73,592,442,400,465,240,66
Lahore,1.533,1.627,147.2,584,602,437,365,340,136
Lima,1.434,1.574,241.57,614,618,545,281,232,61
London,1.295,1.488,619.26,615,628,629,623,625,627
Los Angeles,1.294,1.463,329.49,614,624,616,629,630,427
Luanda,1.518,1.699,138.07,616,623,589,251,182,100
Madrid,1.416,1.557,414.53,609,607,600,583,486,521
Manila,1.399,1.501,271.95,623,586,481,390,296,294
Medan,1.478,1.565,161.89,625,620,591,561,335,43
Mexico City,1.411,1.608,314.42,611,608,624,612,580,455
Miami,1.486,1.589,168.23,546,567,296,171,153,135
Milan,1.348,1.519,562.1,621,611,624,628,627,630
Monterrey,1.413,1.571,218.96,623,622,589,487,324,245
Moscow,1.437,1.564,384.64,600,616,618,619,586,581
Mumbai,1.536,1.654,235.4,588,615,417,283,238,183
Nagoya,1.21,1.397,423.62,629,629,630,627,629,595
Nairobi,1.541,1.669,273.97,603,487,590,511,458,426
Nanjing,1.473,1.576,186.61,609,546,565,505,475,351
New York,1.405,1.561,338.79,536,615,605,624,559,429
Osaka,1.279,1.471,426.54,620,626,623,494,463,494
Paris,1.283,1.465,536.62,619,624,628,627,629,630
Philadelphia,1.331,1.505,411.91,623,598,617,620,622,629
Phoenix,1.328,1.454,318.56,619,625,629,622,592,489
Pune,1.461,1.536,397.43,602,616,618,511,615,373
Qingdao,1.527,1.592,125.23,617,400,174,191,170,180
Quanzhou,1.684,1.717,148.72,168,201,432,390,239,173
Rio de Janeiro,1.674,1.758,337.41,555,476,401,307,225,247
Rome,1.455,1.593,491.23,602,613,598,617,617,400
San Francisco,1.457,1.565,367.15,611,579,344,265,227,228
Sao Paulo,1.405,1.565,490.48,603,618,628,625,621,573
Shanghai,1.351,1.496,215.91,567,623,625,616,558,366
Shenyang,1.414,1.53,125.36,615,620,611,563,507,371
Shenzhen,1.569,1.655,344.57,571,545,595,556,526,369
Singapore,1.518,1.599,152.25,524,525,361,295,223,190
Surabaya,1.574,1.661,232.05,573,579,324,325,285,185
Surat,1.441,1.522,152.92,609,563,400,458,340,187
Suzhou,1.459,1.544,140.48,598,602,549,593,542,416
Sydney,1.475,1.58,316.83,588,611,526,292,222,148
Taipei,1.537,1.672,540.47,597,599,608,523,597,360
Tehran,1.505,1.649,295.42,622,603,603,571,417,300
Tianjin,1.467,1.568,191.99,600,604,586,465,209,86
Tokyo,1.229,1.487,448.21,630,629,630,623,593,495
Toronto,1.304,1.414,229.89,624,593,226,190,189,171
Washington DC,1.312,1.494,566.67,624,606,627,615,624,618
Wuhan,1.558,1.646,211.51,519,570,566,523,536,403
Xiamen,1.57,1.626,176.6,589,482,375,347,305,265
Xian,1.423,1.506,171.11,625,607,611,591,516,562
Yangon,1.469,1.513,162.23,623,624,595,514,340,176
Zhengzhou,1.454,1.532,133.43,613,590,607,533,493,353
\end{filecontents*}
\pgfplotstabletypeset[
	col sep=comma,
	string type,
	every head row/.style={%
		before row={\caption{\bf Data description}\label{si:tab:data_stats}\\ \hline
			\multicolumn{4}{c}{}& \multicolumn{6}{c}{Avg. Number of Valid routes for each radius}\\
		},
		after row=\hline
	},
	every last row/.style={after row=\hline},
	columns/City/.style={column name=City, column type=l},
	columns/short_df/.style={column name=DI (S), column type=l},
	columns/fast_df/.style={column name=DI (F), column type=l},
	columns/via/.style={column name=Avg. waypoint, column type=c},
	]{data_merge_final_1.csv}

	\subsection{Routes as street samples}
	
\begin{figure}[htbp]
		\centering
		\includegraphics[width=0.6\paperwidth]{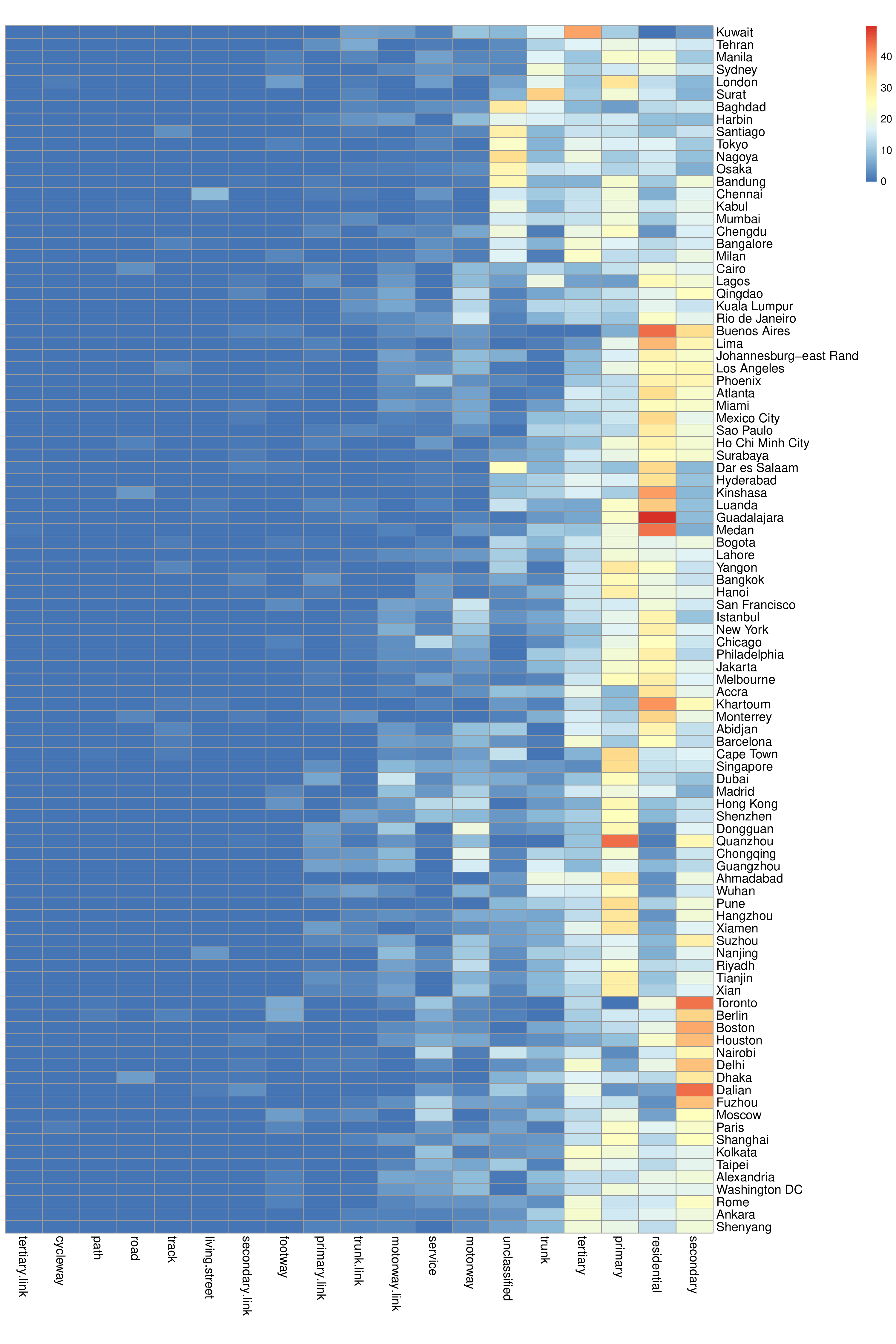}
		\caption[\bf Fraction of road types for shortest routes]{{\bf Fraction of road types for shortest routes}}
		\label{si:fig:road_types_shortest}
	\end{figure}

\begin{figure}[htbp]
		\centering
		\includegraphics[width=0.6\paperwidth]{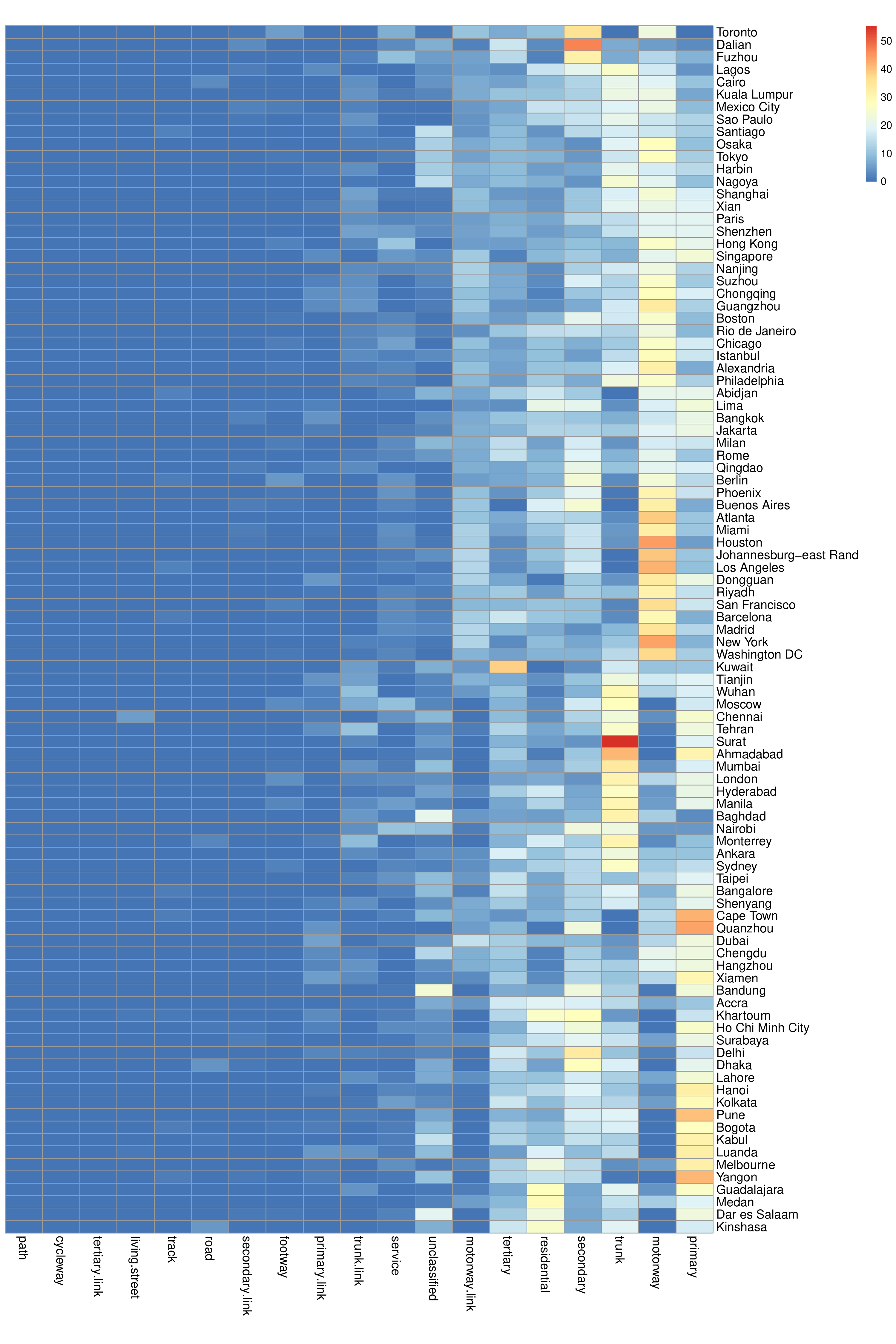}
		\caption[\bf Fraction of road types for fastest routes]{{\bf Fraction of road types for fastest routes}}
		\label{si:fig:road_types_fastest}
	\end{figure}

	If we consider only the subset of the streets within the 30km radius of our analyses, on average, the shortest and fastest routes covered approximately 24\% and 18 \% of the overall streets networks, respectively (Fig.~\ref{si:fig:road_types_distributions}A). However, when we talk about different routing approaches, faster arterial roads and minor residential streets are going to respond for different aspects of the route optimization, and therefore they are expected to cover different samples of a road network. The distribution of road types being sampled by each routing method is depicted in the Fig.\ref{si:fig:road_types_distributions} B\&C. As explained in the main text, arterial roads such as motorway and trunk roads are much more relevant for fastest routes than for shortest routes, reflecting in how frequently they appear in each type of route. 
	Table S2 shows the fraction of the overall streets networks being sampled by the shortest and fastest routes for each city as well as the ratio between the two fractions. Fig.\ref{si:fig:road_types_shortest} \& \ref{si:fig:road_types_fastest} depict the participation of the most frequent road types for each city.
	
	\pgfplotstableset{
		begin table=\begin{longtable},
			end table=\end{longtable},
	}
	\begin{filecontents*}{total_ratio_final.csv}
City, shortest, fastest,ratio
Abidjan,0.149,0.117,0.786
Accra,0.247,0.179,0.722
Ahmadabad,0.285,0.226,0.791
Ankara,0.288,0.191,0.662
Atlanta,0.183,0.141,0.768
Baghdad,0.282,0.183,0.651
Bandung,0.188,0.165,0.874
Bangalore,0.240,0.151,0.627
Bangkok,0.154,0.124,0.808
Barcelona,0.195,0.125,0.641
Berlin,0.186,0.133,0.713
Bogota,0.193,0.125,0.649
Boston,0.248,0.163,0.659
Buenos Aires,0.278,0.192,0.688
Cairo,0.139,0.103,0.738
Cape Town,0.163,0.117,0.718
Chennai,0.177,0.112,0.632
Chicago,0.209,0.126,0.605
Chongqing,0.507,0.486,0.959
Dalian,0.540,0.511,0.947
Dar es Salaam,0.101,0.083,0.821
Delhi,0.092,0.061,0.662
Dhaka,0.185,0.133,0.717
Dongguan,0.181,0.184,1.013
Dubai,0.126,0.100,0.796
Fuzhou,0.367,0.364,0.991
Guadalajara,0.278,0.154,0.554
Guangzhou,0.271,0.248,0.915
Hangzhou,0.343,0.330,0.963
Hanoi,0.354,0.278,0.784
Harbin,0.408,0.374,0.915
Ho Chi Minh City,0.238,0.188,0.791
Houston,0.213,0.129,0.605
Hyderabad,0.097,0.056,0.579
Istanbul,0.204,0.105,0.516
Jakarta,0.178,0.125,0.700
Johannesburg-East Rand,0.222,0.143,0.643
Kabul,0.154,0.118,0.767
Khartoum,0.176,0.112,0.639
Kinshasa,0.135,0.095,0.705
Kolkata,0.197,0.129,0.656
Kuala Lumpur,0.170,0.118,0.697
Lagos,0.309,0.231,0.747
Lahore,0.181,0.136,0.754
Lima,0.231,0.143,0.620
London,0.214,0.129,0.602
Los Angeles,0.250,0.146,0.582
Luanda,0.264,0.177,0.671
Madrid,0.203,0.136,0.669
Manila,0.158,0.108,0.682
Medan,0.321,0.231,0.721
Mexico City,0.222,0.140,0.630
Miami,0.215,0.145,0.676
Milan,0.226,0.131,0.580
Monterrey,0.236,0.136,0.576
Moscow,0.117,0.083,0.706
Mumbai,0.208,0.167,0.802
Nagoya,0.172,0.052,0.305
Nairobi,0.275,0.221,0.802
Nanjing,0.298,0.276,0.927
New York,0.244,0.155,0.634
Osaka,0.153,0.066,0.435
Paris,0.221,0.116,0.527
Philadelphia,0.297,0.189,0.636
Phoenix,0.165,0.099,0.596
Pune,0.368,0.277,0.754
Qingdao,0.404,0.414,1.026
Quanzhou,0.406,0.414,1.019
Rio de Janeiro,0.201,0.150,0.745
Rome,0.289,0.206,0.713
San Francisco,0.213,0.135,0.633
Santiago,0.354,0.275,0.778
Sao Paulo,0.260,0.150,0.575
Shanghai,0.265,0.210,0.792
Shenyang,0.430,0.391,0.910
Shenzhen,0.165,0.146,0.885
Singapore,0.222,0.178,0.802
Surabaya,0.197,0.155,0.785
Surat,0.459,0.391,0.852
Suzhou,0.264,0.259,0.982
Sydney,0.208,0.149,0.715
Taipei,0.230,0.164,0.712
Tehran,0.238,0.148,0.624
Tianjin,0.439,0.397,0.903
Tokyo,0.170,0.058,0.341
Toronto,0.190,0.109,0.577
Washington DC,0.179,0.115,0.645
Wuhan,0.415,0.359,0.866
Xiamen,0.342,0.337,0.986
Xian,0.406,0.355,0.874
Yangon,0.136,0.094,0.695
\end{filecontents*}
\pgfplotstabletypeset[
	col sep=comma,
	string type,
	every head row/.style={%
		before row={\caption{\bf Sample fraction of routes}\label{si:tab:sample_fractions}\\ \hline},
		after row=\hline
	},
	every last row/.style={after row=\hline},
	columns/City/.style={column name=City, column type=l},
	columns/shortest/.style={column name=Shortest route, column type=l},
	columns/fastest/.style={column name=Fastest route, column type=c},
	columns/ratio/.style={column name=Fastest route/Shortest route, column type=c},
	]{total_ratio_final.csv}

\section{Comparison with network centrality measures}
\label{si:sec:network}

In this section, we compare the spatial distribution of inness with different centrality measures of relevance in the context of transport networks. The objective here is to verify to what extent inness can deliver additional relevant information for which traditional measures of centrality do not account. 

Without loss of generality, one can say that the inness of a certain point in the road network of a city is the result of the aggregation of the characteristics of the routes that potentially pass through that point. Indeed, it captures geometric aspects of the network since it is a measure based on the curvature of the roads along a route. It also reflects the structure of the network since it is a metric influenced by topology and network connectivity. Moreover, it also reflects the long-range topographical and geometric relationships since it is not a local measure of centrality but is actually a property of the route. Thus, if a certain point participates in routes with both positive and negative inness, and of equal magnitude, that point is expected to exhibit a null inness. On the other hand, if the inness of the routes crossing a given point is predominantly positive (or negative), we can safely say that that particular point is part of a functional sub-structure of the network with specific geometric characteristics.

We compared the inness with network centrality measures capable of reflecting different facets of the structures, such as topology, locality and geometry. The centrality measures of choice were:
\begin{description}
	\item[Closeness] is a measure of centrality that determines how close a particular node in the network is to all other nodes. Clearly, in the context of spatial networks in a finite area, the nodes with the greatest closeness are those close to the centroid of the network. In the context of spatial networks, closeness is essentially a geometric metric as the actual topology of the network has no importance in determining the centrality of a node. 
	\item[Eccentricity] - the eccentricity of a given node is the largest geodesic distance between that node and any other node in the network. Contrasting with the closeness centrality, eccentricity, by definition, does not reflect the geometric characteristics of the network being mostly a topological metric.
	\item[Degree] centrality is a measure of local connectivity of a node in which the importance of a node is determined by the number of nodes directly connected to it. It is primarily a measure of the local connectivity of the nodes, having no long-range correlations.  
	\item[Betweenness] is a measure that ascribes importance to those nodes lying along the shortest paths between other pairs of nodes such that the more the shortest paths crossing a node, the higher its betweenness centrality will be. It is probably one of the most investigated centrality measures, especially in the context of road networks. 
\end{description}

\begin{figure*}[t!]
	\centering
	\begin{subfigure}[b]{0.31\textwidth}
	\centering
	\includegraphics[height=6cm]{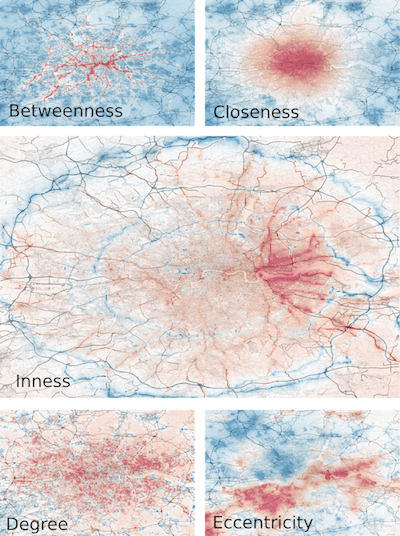}
	\caption{London}
\end{subfigure}
	\begin{subfigure}[b]{0.31\textwidth}
		\centering
		\includegraphics[height=6cm]{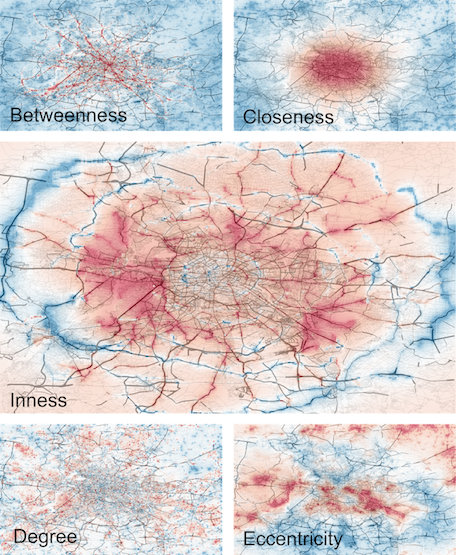}
		\caption{Berlin}
	\end{subfigure}
	\begin{subfigure}[b]{0.31\textwidth}
		\centering
		\includegraphics[height=6cm]{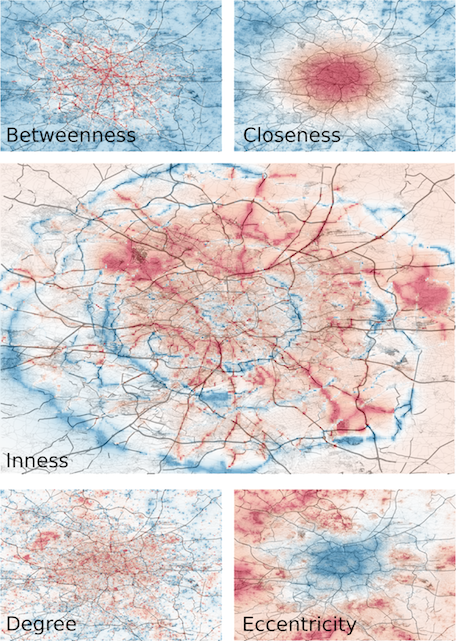}	
		\caption{Paris}
	\end{subfigure}
	\caption[{\bf Comparison of inness and network metrics}] {{\bf Comparison of inness and network metrics} Spatial distribution of inness in comparison with different network centrality measures. Here we show spatial profile of the inness (center panel) compared with the spatial profiles of different centrality measures (bottom and top panels) for three different cities.}
	\label{si:fig:networks}
\end{figure*}

In Fig. \ref{si:fig:networks} the inness profile, along with the four centrality measures for three large European cities. In each panel, values above the average are colored in red, and below that, in blue. The first striking feature we can observe is that the inness profile is very distinct from the ones produced by the centrality measures. For instance, none of the centrality measures managed to capture the concentric circumferential patterns produced by the ring-like structures in the network, manifested as low inness zones.  

Another interesting pattern we can observe is the region of extreme inness profiles, suggesting the presence of large \emph{detour} spots such that routes traveling through those regions have no option but to drive inward. For example, a closer look at the East of London, where routes across the River Thames are being pushed away (low inness in blue) or pulled towards the city center (high inness in red), is a direct consequence of the absence of bridges in that area. Such functional insights from the systems are not possible from the observation of the standard centrality measures in isolation.

\pagebreak

		\section{Types of roads}
		\label{si:sec:types_of_roads}
\begin{figure}[htbp]
		\centering
		\includegraphics[width=0.65\paperwidth]{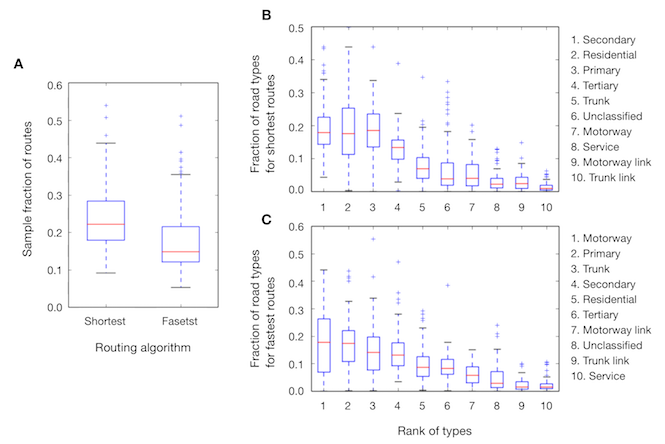}
		\caption[\bf Summary of route information for 92 cities]{{\bf Summary of route information for 92 cities} {\bf A} Sample fraction of routes among entire street networks in each urban area for 92 cities described as boxplots (both shortest and fastest routes). {\bf B\&C} Fraction of road types for sampled routes. {\bf B} shows the shortest routes and {\bf C} represents the fastest routes.}
		\label{si:fig:road_types_distributions}
	\end{figure}

		We summarize the statistics of road types for our sample cities. We also compare the routes samples with the complete road network within the same boundary we used for the inness calculation. The complete road network data was collected from the OpenStreetMap repository using a service\footnote{\url{https://extract.bbbike.org/}}. 
        In Fig. \ref{si:fig:road_types_distributions} we show the distribution of the road types in our routes data for each. See \url{http://wiki.openstreetmap.org/wiki/Key:highway#Values} for detailed information about the road type labels and their meanings. 
	
	\pagebreak

	\section{Summary statistics for average inness}
	
	\begin{figure}[htbp]
		\centering
		\includegraphics[width=0.82\linewidth]{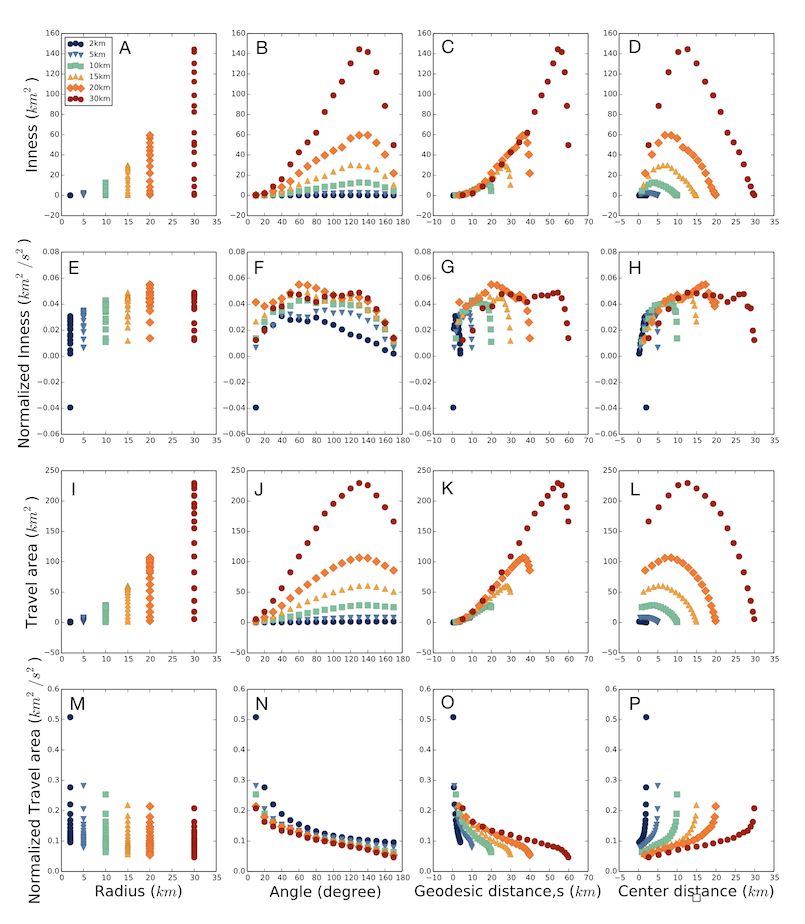}
		\caption[\bf Summary statistics for inness and travel area]{{\bf Summary statistics for inness and travel area}}
		\label{si:fig:summary_inness_stats}
	\end{figure}
	\pagebreak

	\section{Individual cities}
	
	\begin{figure}[htbp]
		\centering
		\includegraphics[width=0.75\paperwidth]{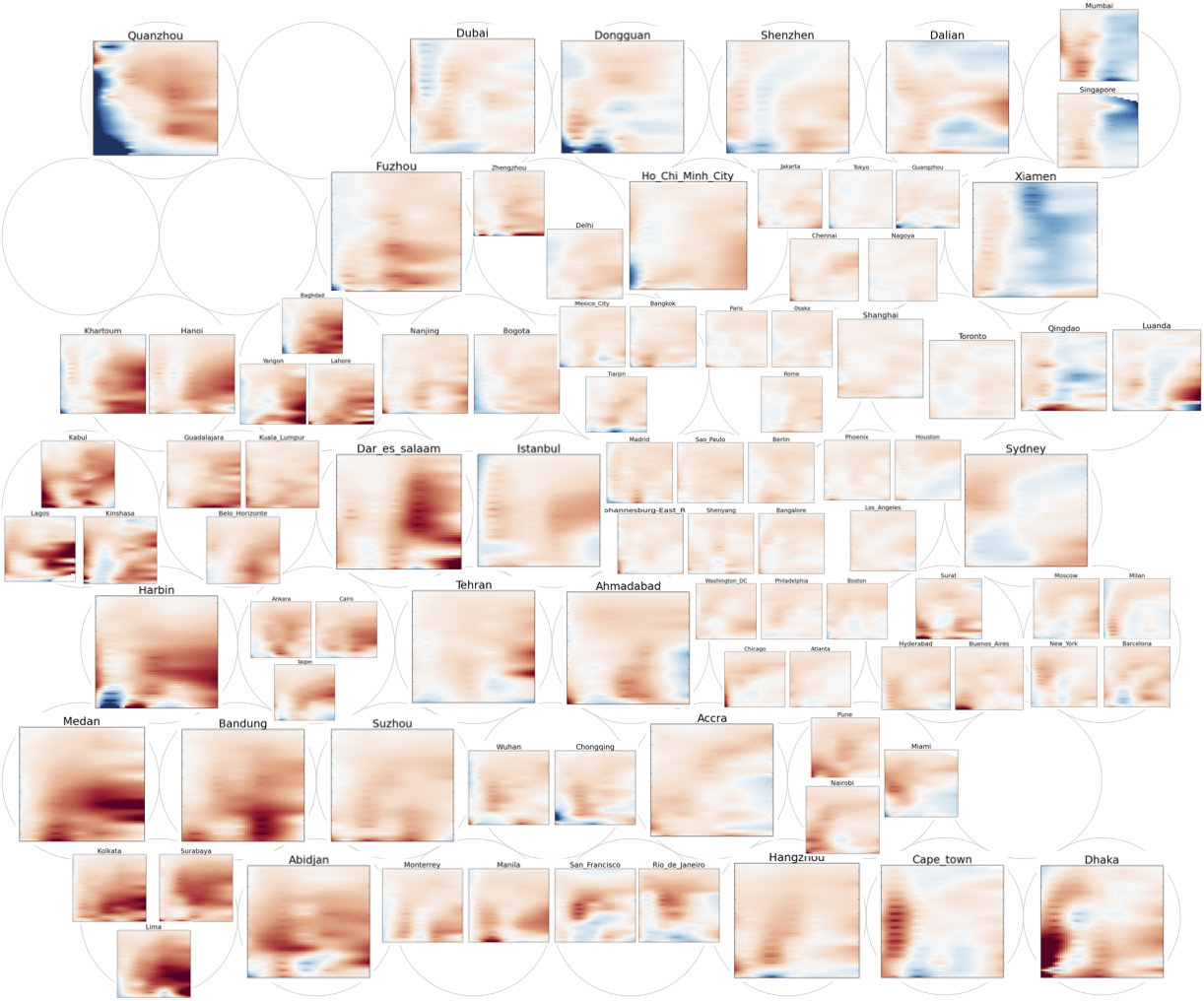}
		\caption[\bf Normalized inness of shortest routes for 89 cities]{{\bf Normalized inness of shortest routes for 89 cities} The normalized inness patterns of shortest routes for individual cities are arranged by its similarity. The cities with similar inness patterns are close to each other and the cities at the both ends are most different to each other. The values range from -0.3 (blue) to 0.3 (red). The cities were clustered using a Self-Organizing Map (SOM) to assign positions in a 2D plane.}
		
		\label{si:fig:som_shortest}
	\end{figure}

	\begin{figure}[htbp]
		\centering
		\includegraphics[width=0.75\paperwidth]{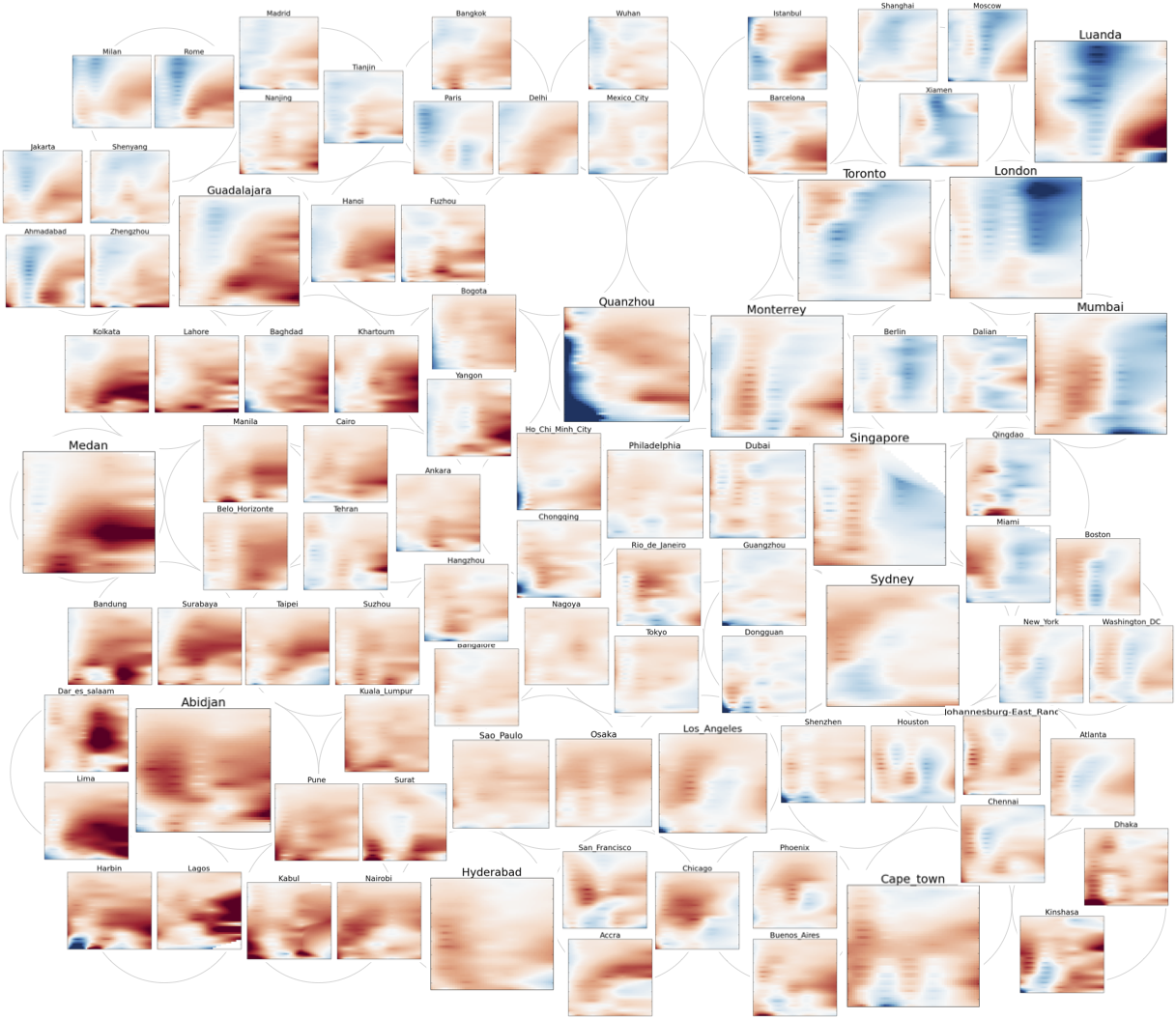}
		\caption[\bf Normalized inness of fastest routes for 89 cities]{{\bf Normalized inness of fastest routes for 89 cities} Method and scale same as in Fig.~\ref{si:fig:som_shortest}}
		\label{si:fig:som_fastest}
	\end{figure}
	\pagebreak

	\subsection{Metric for road structure} 
	\label{si:sec:roads_metrics}
\begin{figure}[htbp]
		\centering
		\includegraphics[width=0.7\linewidth]{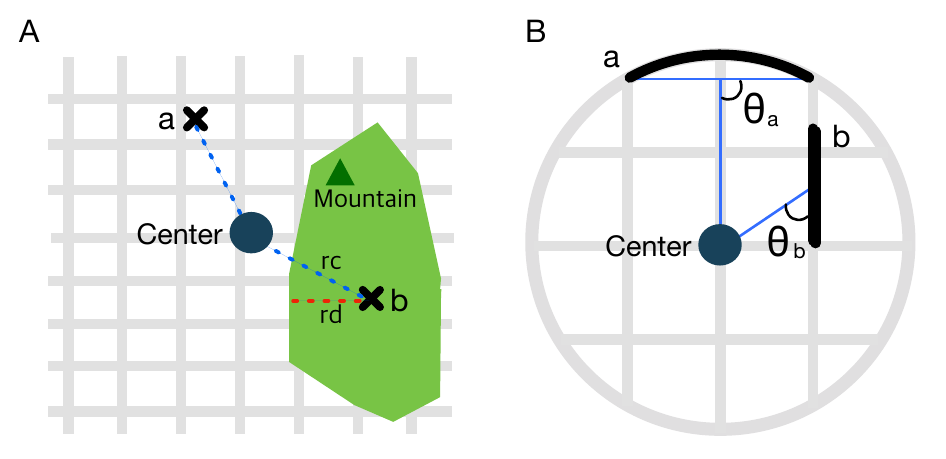}
		\caption[\bf Schematic diagram of road structure metrics] {{\bf Schematic diagram of road structure metrics} {\bf A} Geographical constraints (GC) {\bf B} Peripheral connectivity (PC) }
		\label{si:fig:gc_pc_diagram}
	\end{figure}

\begin{figure}
    	\centering
    	\includegraphics[width=0.8\textwidth]{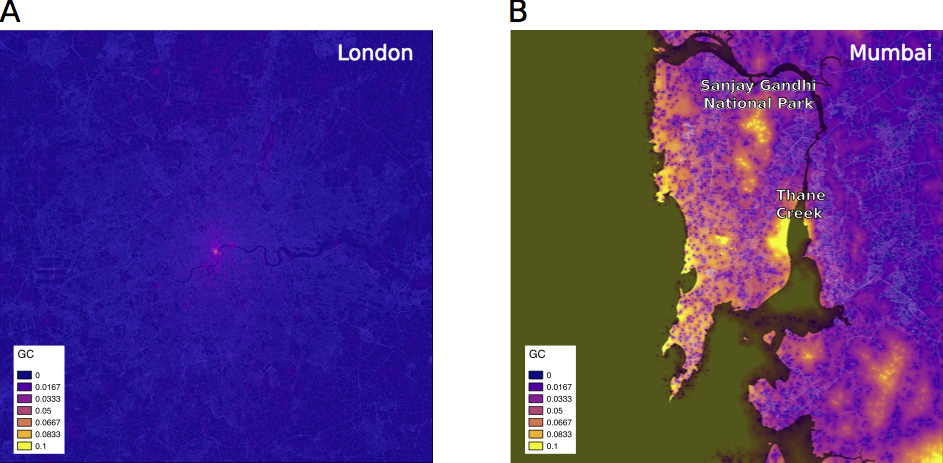}
%
\caption[\bf Examples of the spatial distribution of the geographical constraints (GC)]{{\bf Examples of the spatial distribution of the geographical constraints (GC)} The values of geographical constraints (GC) for 10,000 random points are spatially mapped on two sample cities; {\bf A} London and {\bf B} Mumbai. The same color scheme is applied to both cities with a range from 0 to 0.1. Note that London is one of the cities with low average and standard deviation inness (i.e., LL group) whereas Mumbai is a low average and high standard deviation inness city (i.e., LH group). }
\label{si:fig:geoconstraint}
    \end{figure}

	We suggest three metrics to measure various facet of infrastructural and geographical features. We use same road networks data used and explained in ~\ref{si:sec:types_of_roads}
	\begin{description}
		\item[Road length] is the total length of motorways, trunks, secondary, primary and tertiary roads for each city. 
		
		\item[Level of geographical constraints (GC)] represents the overall fraction of the city that is not covered by the road network due to the presence of \emph{barriers}. Here, we define a barrier as an area of the city that is unaccessible via public roads, being either natural (e.g., forests and mountains) or artificial (e.g., a large industrial or military site). To calculate the GC we  generated 10,000 uniformly distributed points within the same area of our study and computed its relative distance to the nearest point of the road network. More precisely,  GC can be defined as $$GC=rd/rc,$$ where $rd$ is the distance of the point to the closest street segment and $rc$ is the distance to the city center from the random point. For instance, if many random points in a particular area are closer to the center than to a road, GC becomes bigger, suggesting, therefore, the presence of a large barrier close to the center. In the Fig.~\ref{si:fig:gc_pc_diagram}A, the point $a$ is inside the urban area and have road segments nearby while the point $b$ is located in a mountainous area. Although the two points have similar $rc$ (the blue dashed line), $b$ has higher $rd$ (the red dashed line) than $a$ and consequently the GC of $b$ is higher than that of $a$. 
        The term $rc$ accounts for barriers near the city center having a greater impact to routes than barriers of similar area in the periphery. In Fig.~\ref{si:fig:geoconstraint} we show two examples of representative cities with very different GC profiles. For visualization sake, the Arabian Sea is shown underneath a semi-transparent black layer. 
        
        As one can see, London has almost no regions of poor connectivity caused by geographical constraints. Mumbai, on the other hand, has many regions of little to no connectivity due to the presence of geographical constraints such as the large Sanjay Gandhi National Park (the brighter spot in the north-northeast region), and the Thane Creek, the inlet that isolates the city of Mumbai from the Indian mainland.
        
		\item[Peripheral connectivity] represents the average value of all the acute angles of the  \emph{higher-level} peripheral roads, or more precisely, the motorways and trunks beyond a minimum distance from the center, in this case, 10km. These parameter choices are motivated by the reasonable assumption that a road segment with a high angle (or close to 90 degrees) is likely to be part of ring-like structure, which is presumably  used more for connecting peripheries than spoke-like roads. The greater the average angle, the more likely it is that the peripheries are connected, thus acting as a proxy for the presence of circumferential roads. To calculate the angle between the center and a road segment, we draw the shortest line between the center and the middle point of a road segment and measure the angle between the line and the road segment, as depicted in Fig.~\ref{si:fig:gc_pc_diagram}B. 
		
	\end{description}
    
	\begin{figure}[htbp]
		\centering
		\includegraphics[width=0.75\paperwidth]{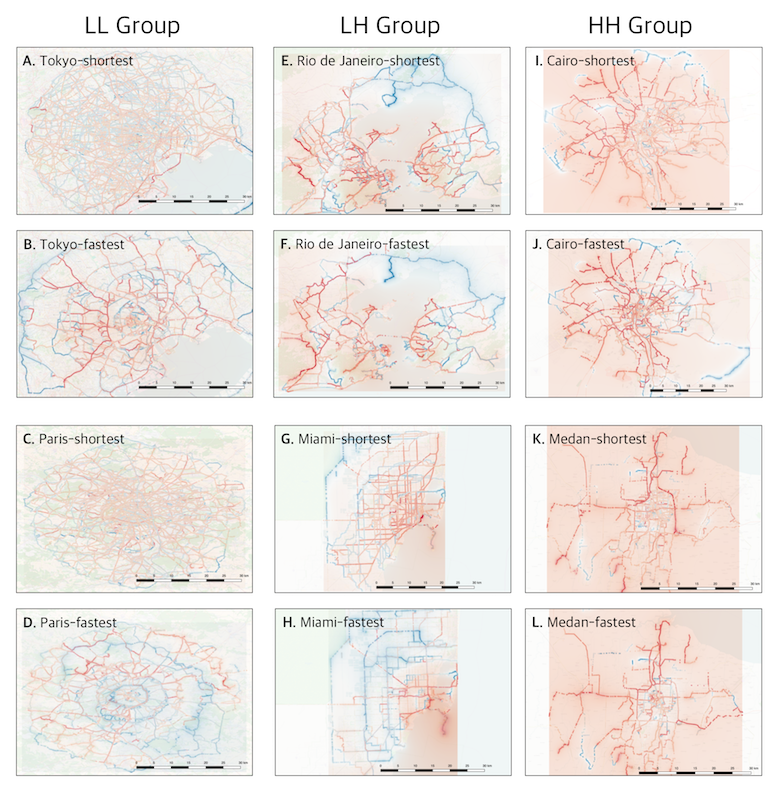}
		\caption[\bf Spatial distribution of cities for each group]{{\bf Spatial distribution of cities for each group.} Examples of cities of the types discussed in Fig. 4. The group LL, LH and HH are classified according to the standard deviation and average values of the inness (LL Group: Low standard deviation and low average (close to zero); LH Group: high standard deviation and low average; HH Group: High standard deviation and high average).}
		\label{si:fig:spatial_distribution_of_cities}
	\end{figure}
	\pagebreak
	
	\subsection{Outlier cities}
	\label{si:sec:outliers}
	Some cities such as Quanzhou, Dongguan, Qingdao, Kinshasa, Harbin, Surat and Kabul exhibit extremely high standard deviation in comparison with other cities (See SI, Section Fig for details on the outliers). For Quanzhou, Dongguan and Qingdao which have relatively low average inness, most part of these cities are shaped by geographical constraints such as the closeness to the coast or being along the path of a river. Just like the geographical constraints influence the shape of the routes of cities in the third category, similar barriers strongly affect these \emph{outlier} cities. For instance, Kinshasa, Harbin, Surat and Kabul basically belong to the second category, i.e., a ``hub and spoke'' structure with strong positive inness signal. However these cities also have negative inness values due to lack of infrastructure (e.g., bridges) connecting different parts of the city across the rivers.
	
\begin{figure}[htbp]
    \centering
    \includegraphics[width=0.7\paperwidth]{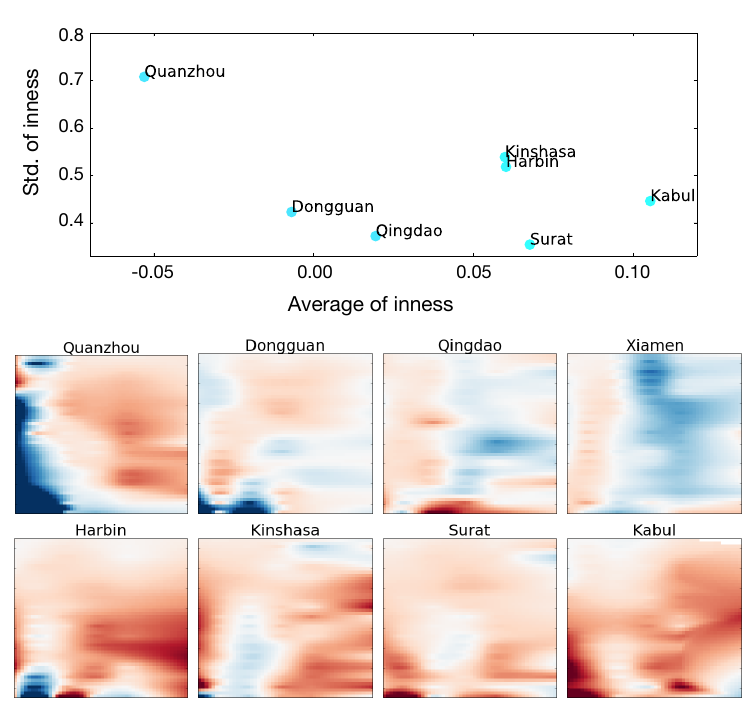}
    \caption[\bf Outlier cities]{{\bf Outlier cities} Inness profiles of cities that do not fall into any of our categories. The values range from -0.3 (blue) to 0.3 (red)}
    \label{si:fig:outliers}
\end{figure}
\pagebreak

		\section{The correlation between shortest and fastest routes}
		When we take into account the inness of the fastest routes, we are indirectly incorporating the influence of second-order structural features of the network such as roadway capacity and speed limits. The analysis of the faster routes, therefore, offers an additional perspective and to a certain point closer to the real operation of that structure.
		
		However, a decidedly more elaborate picture about the structure of the road network can be obtained by means of a quantitative characterization of the geometric similarities and, above all, of the  discrepancies between the shorter and faster routes.
		
		This is because it is only through this comparison that we can verify where and with what magnitude the influence of the path capacity in the geometry of the routes occurs. For example, if for a given pair of origin and destination the shortest and fastest route have discrete inness profiles, this difference is only possible because the segments along the faster route are potentially more temporally efficient.
		
		The correlation between the inness of the shortest and the fastest routes is a measure capable of revealing this difference between the two route types. In fact, those urban systems where the shortest and the fastest routes have little difference are those where any increases in distance are not offset by gains in terms of travel time. From a purely structural perspective this could be said to be an efficient road network such that the fastest routes are also the shorter routes.
		
		Our hypothesis, however, is that it can occur for two main reasons: (1) due to greater homogeneity in terms of road capacity and/or (2) due to low road network capillarity. Therefore, we used three different correlation measures to classify cities according to their similarity profiles between the shortest and fastest routes. 
		
		\subsection{Classifying cities based on their Inness profiles}
		\label{si:sec:clustering}
		Although we are not claiming that the cities can be \emph{naturally} classified into different discrete groups, here we show that the correlation between the inness of shortest ($I_{s}$) and fastest ($I_{f}$) routes can be used as a metric to classify cities. Thus, we computed three correlation measures, namely Pearson correlation, Spearman's rank order correlation and Kendall rank correlation. The measures were computed comparing the inness of the average shortest and fastest routes for each radius/angle value.
		
		For each of the $\mathcal{N}\leq 36$ routes with radius $r$ and angular distance $\theta$ we computed the average $I_{S}$ and $I_{F}$, with the inequality being due to the existence of unfeasible paths for certain OD pairs, and measured the correlation coefficients between the two inness arrays. The rationale to use three correlation metrics is that this way we can characterize the said dissimilarities in a higher dimensionality space, accounting not only for the absolute values of the inness but also for the ranks of the $(r,\theta)$ pairs in terms of their inness . 
		
\begin{figure*}[t!]
\centering
\includegraphics[width=0.9\linewidth]{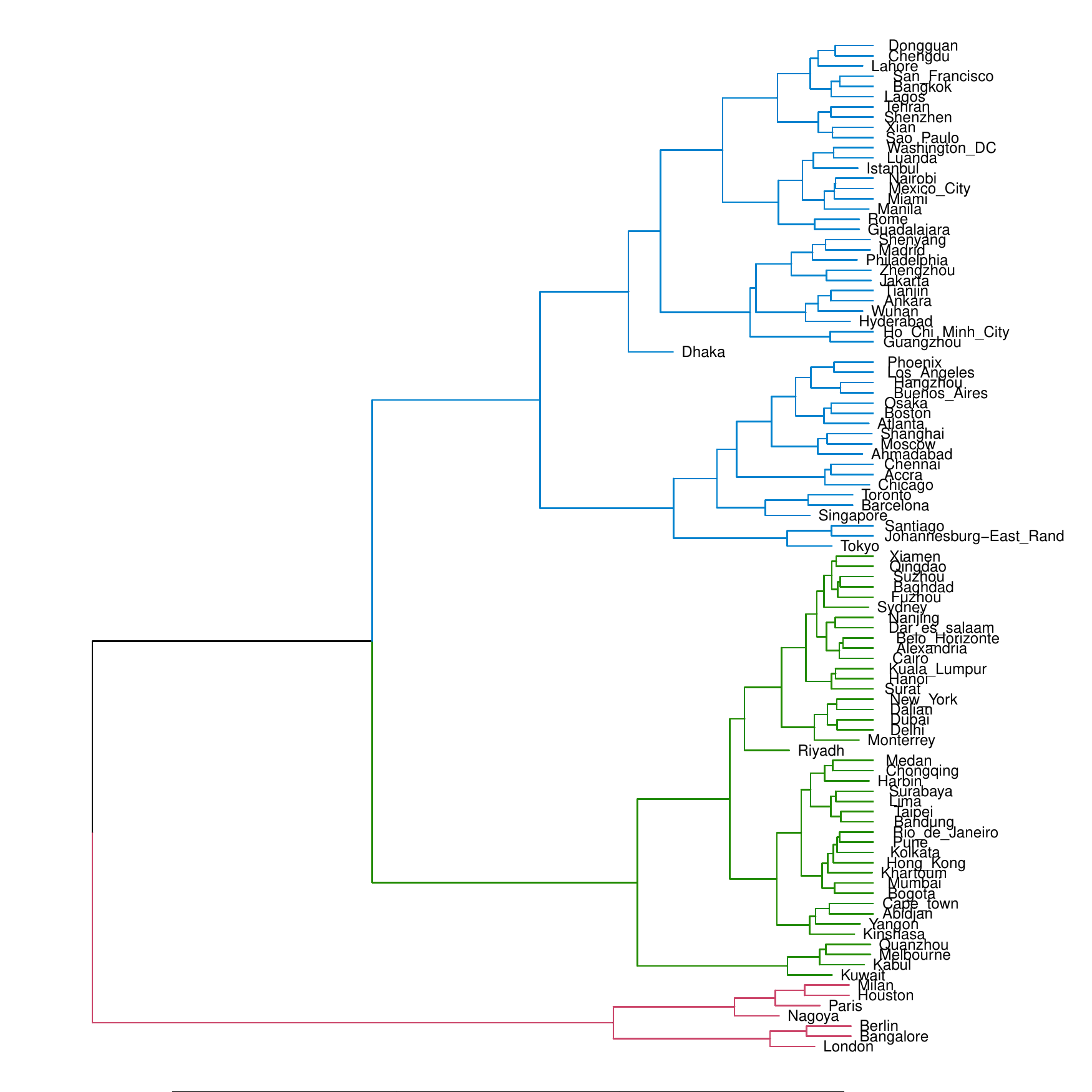}
\caption[{\bf Hierarchical clustering of cities based on three correlation measures}]{{\bf Hierarchical clustering of cities based on three correlation measures} The colors illustrate a 3-clusters partitioning.  
}
\label{si:fig:clusters}
\end{figure*}

		We then applied a hierarchical clustering method to produce \emph{a} partitioning of the cities based on their similarities in terms of their fastest and shortest inness profiles.  The method is a standard complete linkage clustering method in which the maximum possible distance between points belonging to different groups is sought. Fig. \ref{si:fig:clusters} shows the dendrogram of the partitioning. 
		
\begin{figure*}[t!]
			\centering
			\includegraphics[width=0.7\linewidth]{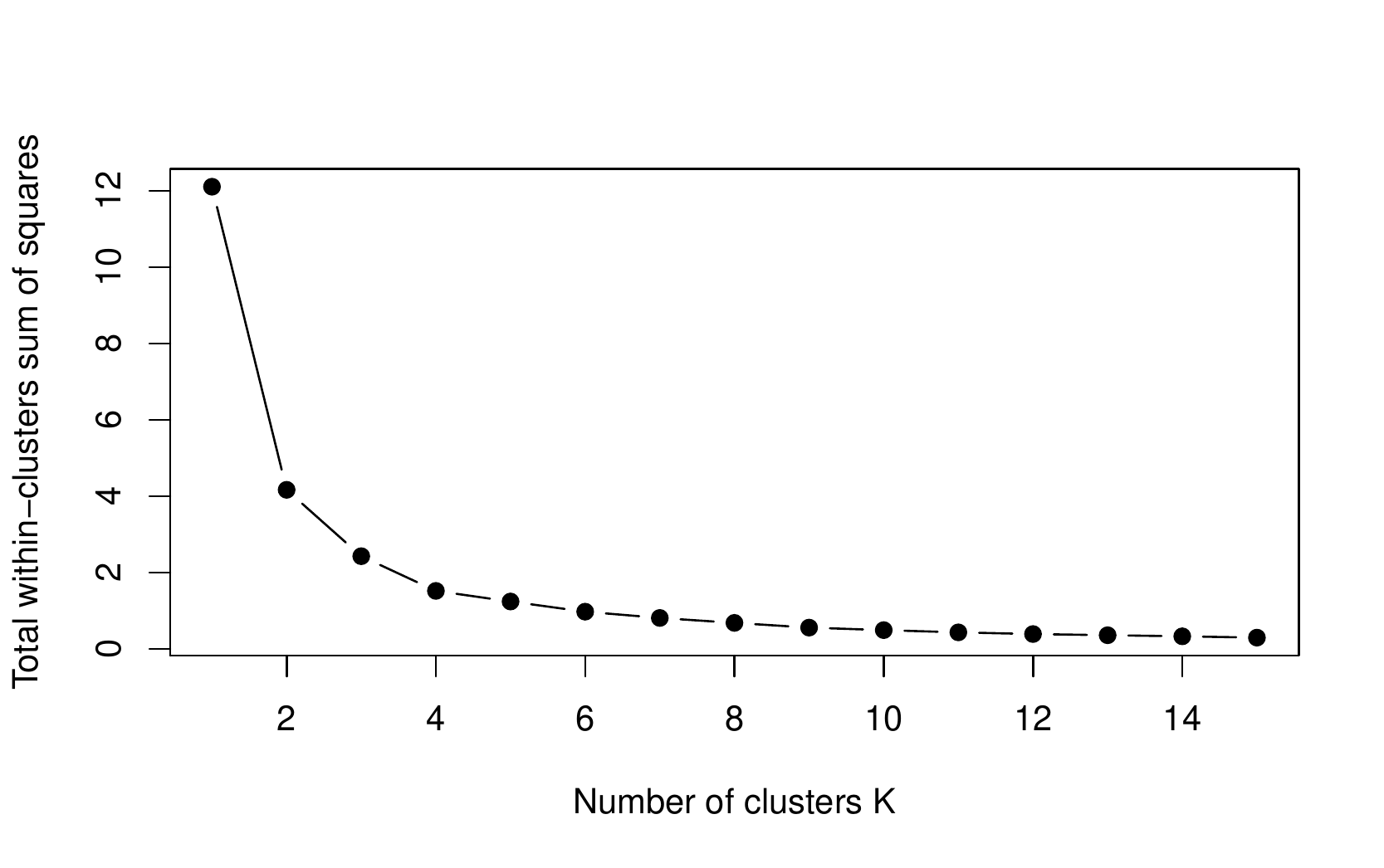}
			\caption[{\bf Within-clusters sum of squared deviations (WCSS) as a function of the number of clusters $k$}]{{\bf Within-clusters sum of squared deviations (WCSS) as a function of the number of clusters $k$} Most of the variance can be explained by three clusters and only very little variance is explained by increasing the number of clusters from 4 to 5, suggesting that the best partition would be one with $k=3$ or $k=4$.  }
			\label{si:fig:numofclusters}
	\end{figure*}

		Next we computed the within-clusters sum of squared deviations (WCSS) to quantify how much of the variance could be explained by partitioning the cities into$k$ clusters. Clearly, a perfect partitioning would be one in which each cluster contains one single city. Fig. \ref{si:fig:numofclusters} shows the WCSS as we increase the number of clusters. As we can see, most of the variance can be explained by three clusters and only very little variance is explained by increasing the number of clusters from 4 to 5, suggesting that the best partition would be one with $k=3$ or $k=4$.

	\pagebreak

\subsection{Socio-economic indicators}
\label{si:sec:indicators}
As we presented in the main manuscript, the (dis)similarities between the inness profiles produced by the shortest and fastest routes are often rooted on the level of development of the road infrastructure, which in turn is driven by the socio-economic development of the cities. We then explored the correlation between the $I_{s}$ and $I_{f}$ with three relevant indicators that could reflect the said stages of developments, namely the \emph{productivity index (PI)}, the \emph{infrastructure development index (IDI)} and the \emph{GDP per capita} of the cities. The first two indexes (PI and IDI) are part of the \emph{City Prosperity Index}, to date, the most comprehensive measure of the development of a city, developed by the United Nations program for human settlements  (UN-Habitat). Each one of the six CPI indexes (including the PI and IDI) is defined in terms of an array of other indicators such as household income, economic specialization and housing infrastructure. The decision to employ the PI and IDI is motivated by the fact that these are the indexes more closely related to the structural development of the cities than other ones. For more details on the CPI indexes we refer the interested reader to the UN-Habitat Methodology and Metadata report \footnote{\url{http://cpi.unhabitat.org/sites/default/files/resources/CPI\%20METADATA.2016.pdf}}.


The third indicator we used, i.e., the GDP per-capita of the cities,  is based on the GDP@Risk estimate, a projected GDP of the cities based on the World City Risk Index ---  a risk-assessment metric developed by the Cambridge Centre for Risk Studies and published on the Lloyd's City Risk Index. More precisely, the index is a projection from 2015-2025 of the GDP accounting for different risk factors for the 301 world's major cities. More detailed information on the methodology can be found in the report `World City Risk 2025: Part 2 Methodology' \footnote {\url{http://cambridgeriskframework.com/wcr}}.

The decision to use a \emph{projected} GDP -- instead of the official estimated nominal GDPs officially published by the governments -- is justified by three main reasons. The nominal GDP of a city is subject to some volatility due to many internal and external factors, contrasting with the transportation infrastructure of a city that tend to evolve over longer periods of time. Moreover, there are a lot of methodological variation in the way the nominal GDPs are estimated, especially for non-OECD cities. Additionally, the most recent data of the official  GDPs  does not necessarily correspond to the same period for different cities. 

On the other hand, the projected GDP of the cities is a standardized metric based on the same scientific methodology for all the cities accounting for many factors of internal and external origin, from present infrastructure to potential natural disasters. Moreover, the GDP projection can reflect with a reasonable precision the \emph{potentialities} of growth for a city, in which the level of development of the infrastructure plays a major role.  

	\subsection{New York}
	\label{si:sec:new_york}
	Contrasting with other large developed urban cities (type I), New York exhibited similar inness pattern between the shortest routes and fastest routes. The reason for such phenomena can be related to the geography of the motorways in the New York metropolitan area. Unlike other cities, New York does not have strong ring-like motorway structure in its periphery, which often are the preferred structures when it comes to congestion reduction and travel-times optimization. Instead, it has many radial and grid-like motorways, which has a limited effect on the inness patterns, as shown in the spatial distribution of fastest routes. Such particularity of the motorways of New York gives it unique inness characteristics, although further investigations regarding other factors (e.g., socioeconomic characteristics) is necessary.
	
	\begin{figure}[htbp]
		\centering
		\includegraphics[width=0.75\paperwidth]{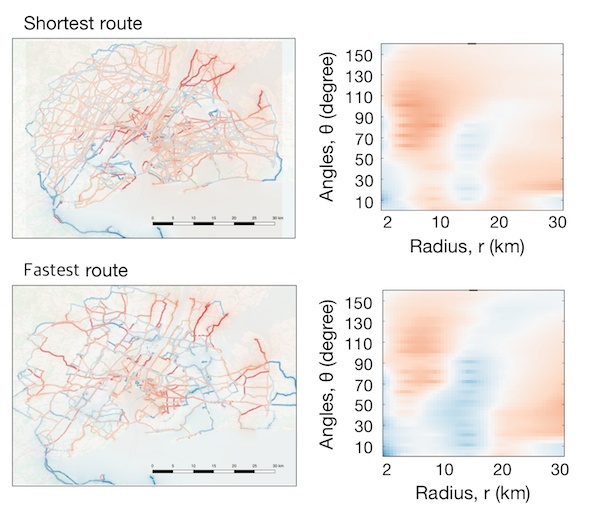}
		\caption[\bf Inness pattern and spatial distribution of New York]{{\bf Inness pattern and spatial distribution of New York.}}
		\label{si:fig:new_york}
	\end{figure}

%
	
	\clearpage



\end{document}